\documentstyle[floats,preprint,oldlfont,aps,epsf,here]{revtex}
\begin{document}

\title{How to derive a  protein folding potential? \\
A new approach to the old problem.}  
\author{L. A. Mirny, E. I. Shakhnovich} 
\address{Harvard University, Department of
Chemistry\\ 12 Oxford Street, Cambridge MA 02138} \date{\today}
\maketitle
\begin{center}
\vspace{10pt}

\vspace{10pt}

Running title: Protein folding potential.

\vspace{20pt}

Correspondence to E.Shakhnovich \\ Dept of Chemistry Harvard
University\\ 12 Oxford Street Cambridge, MA 02138\\ tel
617-495-4130 \\ 
FAX 617-496-5948 \\ 
E-mail: eugene@diamond.harvard.edu

\end{center}

\newpage

\vspace{20pt}
\section*{Abstract}
In this paper we introduce a novel method of 
deriving a pairwise potential for protein folding. The
potential is obtained by optimization procedure, which simultaneously
maximizes the energy gap for {\it all} proteins in the database. 

To
test our method and compare it with other knowledge-based 
approaches to derive potentials, we
use simple lattice model. 
In the framework of the
lattice model we build a database 
of model proteins by a) picking randomly 200 lattice chain conformations; 
b) designing sequences which fold into these
structures with some arbitrary ``true'' potential; 
c) use this database
for extracting a potential; d) fold model proteins using
the extracted potential. This test on the model system showed that
our procedure is able to recover the potential with correlation
$r \approx 91\%$ with the ``true'' one and we were able to fold
all model structures using the 
recovered potential. Other statistical 
knowledge-based approaches were tested 
using lattice models
and the results indicate that they 
also can recover the ``true'' potential
with high degree of accuracy.

When applied to real protein structures 
with energy function taken in contact 
pairwise approximation, our potential 
scored somewhat better than
existing ones. However,  the discrimination 
of the native structure from
decoys is still not strong enough to make the potential useful
for {\em ab initio} folding. We argue that  
more detail of protein structure and energetics should be
taken into account to achieve better energy gaps.
The suggested method is general  
to allow to systematically derive parameters
for more sophisticated energy function.  
The internal control of
validity of the potential derived by our method, 
is converegency to a unique solution
upon addition of new proteins to the database. 

\pagebreak

\section*{Introduction}
The problem how to determine correct energetics is paramount
to the complete solution of the protein folding problem.

Two avenues to determine energy functions (force-fields) for proteins
have been pursued. First, is more or less rigorous or semi-empirical
classical or quantum mechanical calculations to determine, from the 
first principles,
and/or fitting to spectroscopic experimental data, the forces
acting between aminoacids in vacuum or 
in solution (Vasques {\em et al}, 1994).
This approach is rigorous but 
it encounters formidable computational
difficulties. Most importantly, it can be realized only within the
framework of detailed, atomistic description of aminoacids. However,
detailed atom resolution models of proteins are 
not feasible for folding simulations
due to obvious computational difficulties.

An alternative, more practical, approach is to introduce simplified, coarse-grained
models of proteins where aminoacids are represented in a simplified
way, as one or few interacting centers which may have
some internal degrees
of freedom as well but which are generally much simpler
than real aminoacids (Levitt, 1976; 
Ueda {\em et al}, 1978; Miyazawa and Jernigan, 1985; 
Wilson and Doniach, 1989; 
Skolnick and Kolinsky 1990; 
Shakhnovich {\em et al}. 1991). 
Such models are more
tractable computationally both in 
threading approaches (Finkelstein and Reva, 1991; 
Jones {\em et al}, 1992) and 
and {\it ab initio}  simulations 
(Kolinsky and Skolnick, 1993, 1994). 
However, the serious
problem with simplified 
representation of proteins,
is how to describe protein 
energetics at the coarse
grained level of structure description. 
In particular, what ``force-fields'' 
should act between these simplified 
interacting centers,
which are still identified with natural aminoacids, 
such that native structures,
for these model proteins, 
still correspond to pronounced energy minima for their
respective sequences? An approach to address this problem
was proposed by Tanaka and Scheraga (1976) and
was developed  in the seminal contribution 
of Miyazawa and Jernigan (MJ)
(1985). 
The MJ method is based on  statistical analysis
of protein structures  and determination of frequencies
of contacts, defined in the realm 
of simplified protein representation, e.g.,
as two $C_{\alpha}$ atoms being 
closer to each other than specified cutoff distance.
Frequencies of individual aminoacid 
contacts were derived and compared
with what one should expect at
random mixing, and interaction with solvent
was accounted for as well. Then quasichemical approximation
was employed which related these properly normalized frequencies
with ``potentials'' via the relation:
\[
u_{ij}=-T \ln f_{ij}
\]
where i and j denote aminoacid types; $f_{ij}$ 
are normalized
frequencies of contacts between them determined from the database
of existing structures. 
The definition of energy scale denoted as ``Temperature''
T in the quasichemical approximation of MJ is a delicate problem. It 
was addressed in a 
recent
work by Finkelstein {\em et al} (1993,1996) who also showed that
quasichemical approximation may be a reasonable one 
under the assumption that
protein sequences are random. In the recent work Mirny and Domany (1996) 
showed that quasichemical approximation is also valid if
conatct energies are independent and unformly disributed. 
The subsequent development of 
the knowledge-based approach based on quasichemical approximation
included efforts to incorporate distance-dependent forces
Sippl (1990), better representation of aminoacid geometry and
approximation od multible-body interactions 
(Kolinsky and Skolnick, 1993,1994), 
dihedral angles (Kolaskarand and Prashanth, 1979; 
Nishikawa and Matsuo, 1993;
Rooman {\em et al}, 1992; DeWitte and Shakhnovich, 1994), 
better treatment of solvent-protein 
interactions (Levitt and Hinds, 1995). A detailed analysis
of features of knowledge-based potentials and examples of their
successful and unsuccessful application is given by
Kocher {\em et al} (1994); the approaches to derive potentials
from quasichemical approximation, especially the 
most difficult issue of reference state are  discussed by
(Godzik {\em et al}, 1995).

Real potential is believed to discriminate the native structure by
making its energy much lower than energy of all other
conformations, i.e. it provides stability of the native structure. 
Protein sequences should also fold fast to their respective
native conformations. It was shown,
for simple models of proteins, that these two conditions - thermodynamic
stability and kinetic accessibility - are met when
the native state is a {\em pronounced} energy minimum for the
native sequence, compared to the set of misfolded
conformations (Sali {\em et al}, 1994; 
Shakhnovich, 1994; Gutin {\em et al}, 1995). Therefore it is reasonable
to suggest that  the essential property of the
correct, ``true'', 
folding potential, is that the energy of a  native sequence folded
into its respective native conformation 
should be much lower than the energy of
this sequence in all alternative conformations.

An  approach to derivation of protein folding 
potentials which takes this requirement explicitly into account 
was proposed by Goldstein {\em et al} (1992) (GSW) and by Maiorov
and Crippen (1992). Goldstein {\em et al} 
maximized the quantity $T_f/T_g$, equivalent to maximization of the
energy gap between the native state and bulk of decoys.  
They showed that 
for each individual protein the problem
of potential optimization
has a simple analytical solution; however 
their approach encountered a serious problem
of how to average over different structures in the database.
Indeed, a potential optimized for one protein is
not necessarily (an in fact never!) optimal for another protein,
while the goal is to find a potential which 
is good, or optimal, simultaneously
for many proteins. GSW found an {\em ad hoc} procedure 
of averaging over protein database 
which gave  
best results in their tests.

In this work we suggest a
systematic method to find a potential 
which delivers pronounced energy minimum to {\em all}
proteins in their native conformations 
and hence should provide fast folding
and stability of model proteins. 
The method is general
and is not limited to any form 
of potential or any model of a protein. Another
important feature of this approach is that it has internal
criterion of self-consistency: when the derived potential
does not change significantly upon addition of new
proteins to database it corresponds to meaningful, nontrivial energetics.

The proposed new method of potential derivation should
be tested and compared with existing approaches. How can it be done? 
The serious problem with testing parameters derived by 
any approach 
is the lack of objective rigorous criterion of success
because ``true'' potentials are not known (and they are not likely
to exist at all since real proteins differ from their
simplified representation). A reasonable 
criterion can be that the derived potentials
are useful
for fold recognition  and
{\it ab initio} folding. The results of numerous tests by many groups 
(see the comprehensive analysis in the special issue of
Proteins, (Lattman, 1995))
show that while existing knowledge-based potentials often 
do a decent
job in fold recognition, they are not sufficiently
accurate
for {\it ab initio} folding. A strong evidence 
that the ``bottleneck'' in {\it ab
initio} folding 
is in the energy function rather than in search strategy is that
{\it ab initio} procedures fail because decoys with energy, lower than
energy of expected native conformations are found in test cases
(Covell, 1994; Elofson, 1995), while in inverse folding tests the native structure
is in most cases (though not always) has 
lower energy than decoys (Wodak and Rooman, 1993, Lemer {\em et al}, 1995; Miyazawa and Jernigan, 1996). 

The best way to assess different procedures of derivation of potentials
is to use, as a test case, 
a model system where the correct form of the Hamiltonian (say, pairwise
contact potential)
is given, and the ``true'' potential is known. Different
procedures to extract potentials can be applied, 
and 
then the  ``true'' and derived potentials can be compared.
Further, the derived potentials can be used, in a model system, 
for 
threading or {\it ab initio}
folding to compare its performance 
with  such of ``true'' potentials and thus to close the circle.

Such a comprehensive analysis is possible in the realm of lattice
models. The step in this direction was made by Thomas and Dill (1996);
These authors considered 2-dimensional short lattice chains
composed of  monomers of two types.

In this paper we test our procedure of derivation 
of potentials as well as other approaches using 3-dimensional
lattice model proteins composed of  ``aminoacids'' of 20 types. 
Sequence design algorithm has been developed recently
which generates sequences which have specified
relative energy (Z-score, Bowie {\em et al}, 1991)
in a given conformation
Shakhnovich and Gutin, 1993a,b;
Abkevich {\em et al}, 1995a).  
. 
This allows us to carry out the following rigorous procedure
of testing our and alternative approaches to derive
parameters:

a) Select at random a number of lattice conformations
to serve as native ones.

b) Using some potential (``true'' potential 
for the model), design
sequences to have selected 
conformations as native ones, thus creating
a model ``protein data bank''. 

c) Using different procedures  extract ``knowledge-based potentials''

d) Compare derived potentials with the ``true'' one. Test the performance
of the 
derived potentials in {\it ab initio} folding simulations and threading,
using the database of model proteins which has not been used for
the derivation of potentials.

This approach to test the potentials 
allows to address, in a systematic
way,  and for different procedures of parameter derivation,
an important issue of what database size is sufficient to 
successfully derive parameters.
Further, we can create ``databanks'' of model proteins
of different stability. It makes it possible to address the
issue of how well optimized protein sequences should be
to allow successful derivation of parameters.

The program described in a-d  (except {\it ab initio} folding tests) 
can be straightforwardly carried out
not only for lattice model proteins but for real proteins. Indeed,
with any set of parameters, we
can design sequences for protein conformations, and evaluate 
the ``ideal'' value of Z-score for a given model Hamiltonian.
This is to be compared with scores for native sequences
with the same parameters. The comparison will shed light not only
on advantages and pitfalls of parameter derivation procedure
but also on the most important issue of what models are better
for prediction of protein conformations.

In subsequent 
chapters of this paper we will 
carry out this program.

\section*{Methods} 
\subsection*{Derivation of potential}

Energy function assigns the value of the energy to a given conformation
for a given amino acid sequence.
\begin{equation}
   E=E({\text Sequence},{\text Conformation},{\bf U})
\label{eq:E0}
\end{equation}

Where {\bf U} is the set of parameters of potential to be derived from
known native structures of proteins.

We use  $Z$ score  as a measure of how pronounced is the energy minimum corresponding to
the native conformations (with respect to a set of alternative
conformations)  (Bowie {\em et al}, 1991):

\begin{equation}
   Z= \frac{ E_N - \langle E \rangle _{conf} }{ \sigma_{conf} (E)}
\label{eq:Z}
\end{equation}

which is the deviation of the energy of the native conformation from
 average energy of alternative conformations measured in
units of standard deviation.  Average energy $ \langle E \rangle
$ and variance $ \sigma (E) $ are computed over a set of
alternative conformations (see below). Absolute value of $Z$
score is the natural measure of the energy gap. 

Our goal is to find a potential {\bf U} which minimizes $Z$ scores
(maximizes the energy gap) simultaneously for all proteins in the
dataset. This is achieved by building a target function, which is an
appropriate combination of
individual $Z$ scores and then optimizing this function with respect
to {\bf U}. One should be careful about choosing a combination
of $Z$ scores to optimize. If the target function to be
optimized is chosen naively, for example sum of $Z$ scores,
then low values of the target function can be obtained if $Z$
is small enough for some proteins and large for all others.  To
avoid this kind of a problem one has to minimize $max_{m} (
Z^{(m)} )$, which is however, very difficult to deal with
because of its discontinuity.

We chose harmonic mean of $Z^{(m)}$ scores as a function to be
minimized.

\begin{equation}
\langle Z \rangle _{harm} = \frac{M}{ \sum_{m=1}^{M}\frac{1}{Z_m}}
\label{eq:Zavr}
\end{equation}

In fact, harmonic mean is a smooth approximation of
$max_{m}(Z^{(m)})$ since terms with the smallest absolute
value of $Z^{(m)}$ scores contribute most to the harmonic mean.

To maximize the energy gap for all proteins in a dataset we
search for a potential {\bf U} which maximizes the value of
function $F({\text {\bf U}})=  - \langle Z \rangle _{harm}$. 
The value $F$ is directly related to the energy gap so in future 
we will (somewhat semantically frivolously) refer to as energy gap,
understanding though that it is not exactly identical to it,
but there is a monotonic dependence bwteen these two quantities.

We also
apply some constraints on the potential {\bf U}:

\begin{equation}
\langle U \rangle = 0
\label{eq:const1}
\end{equation}

\begin{equation}
\sigma ^2 (U)= \langle ( \langle U \rangle -U)^2 \rangle = 1
\label{eq:const2}
\end{equation}

The first constraint sets average interaction between amino
acids to zero, i.e. eliminates non-specific attraction/repulsion
between amino acids.  The role of the second constraint is to set
dispersion of interaction energies to one. If energy is a linear
function of parameters, multiplication of $U$ by an arbitrary
constant does not change values of $Z$ score and by setting
$ \sigma(U)=1$ we chose units of energy.

Potential $U$ is obtained by maximization of $F({\text {\bf U}})$
using a procedure for non-linear optimization.  The potential obtained
in this way is (by procedure) the one which provides the largest
energy gaps simultaneously to all proteins in the dataset as far
as $\langle Z \rangle _{harm}$ is an accurate approximation
to $max_{m}(Z^{(m)})$.

The very important part of the method is choice of alternative
conformations to compute $Z$ scores. In general one has to use the
same set of conformations for sampling and for computing $Z$
scores. For example, to optimize potential for threading one has to
compute $Z$ scores using a set of alternative conformations obtained by
threading a sequence through representative set of protein
structures. This procedure is not computationally expensive since for
threading the set of alternative conformations does not depend on the
potential used.  To find a potential one needs to generate a set of
alternative conformations, and then, use this set to compute
individual $Z$ scores. However, when dynamic sampling
techniques (Monte Carlo, Molecular Dynamics, Growth procedures
etc),  are used for {\it ab initio} folding,
the set of alternative conformations is not known in
advance and, what is more important, depends on the potential
applied. In this case one has to make some assumptions about the
ensemble of alternative conformations which will allow to compute
average energy $\langle E \rangle _{conf}$ and variance of energy
$ \sigma  _{conf} (E) $ over the of ensemble of alternative conformations.
Here we show how to optimize pairwise potential for {\it ab
initio} folding of a simple model and for threading.

\subsection*{Derivation of parameters for pairwise potential.}

\subsubsection*{Pairwise potential}
Here we consider pairwise contact potential, i.e.  the energy of a
conformation is a sum of energies of pairwise contacts between
monomers, which are not nearest neighbors in sequence:

\begin{equation}
E(\mbox{\boldmath $\xi,\Delta$ })=\sum_{1\leq i < j\leq N} U(\xi_i,\xi_j)\Delta_{ij}
\label{eq:ENER}
\end{equation}

where $\Delta_{ij}=1$ if monomers $i$ and $j$ are in contact and
$\Delta_{ij}=0$ otherwise. 
Various definitions of contacts can be used (Kocher, 1994).
$\xi_i$ defines the type of amino acid
residue in position $i$. Potential is given by $U$ matrix, where
$U(\xi,\eta)$ is energy of a contact between amino acids of types
$\xi$ and $\eta$.

\subsubsection*{Optimization of potential}
The optimization of $F({\text {\bf U}}) = - \langle Z \rangle _{harm}$
is performed by Metropolis Monte Carlo procedure in space of
potentials, i.e. at each step a cell $U(\xi, \eta)$ of the matrix {\bf
U} is chosen randomly and a small random number $r \in [-0.1,0.1]$ is
added to $U(\xi, \eta)$. This change is accepted if it increases
$F({\text {\bf U}})$ and rejected with probability $1- \exp ( - \frac{
\delta F}{T_{opt}})$ if it decreases $ F({\text {\bf U}})$. $T_{opt}$
is the ``temperature'' of optimization. Optimization of potential
starts from a completely random potential and stops when target
function changes less than on $\epsilon=0.01$ for last 20000 steps.


\subsubsection*{Computation of $Z$ score}
For a given sequence \mbox{\boldmath $\xi$}, potential {\bf U} and
generated set of alternative conformations \mbox{\boldmath
$\Delta^{(k)}$} $ k=1,..K$ one can compute $Z$ of the native conformation
\mbox{\boldmath $\Delta ^N$ }:

\begin{equation}
Z( \mbox{\boldmath $\xi$}, \mbox{\boldmath $\Delta ^N$})=
 \frac{ E(\mbox{\boldmath $\xi$}, \mbox{\boldmath $\Delta ^N$})- 
\langle E(\mbox{\boldmath $\xi$}, \mbox{\boldmath $\Delta^{(k)}$} ) \rangle _k }{
\sigma (E(\mbox{\boldmath $\xi$}, \mbox{\boldmath $\Delta^{(k)}$} )) _k}
\end{equation}

where index $k$ denotes averaging over alternative conformations.

Instead of computing energy of a sequence in each alternative conformation 
each time we need $Z$, we compute some average quantities for the set of conformations
and then use them to estimate $Z$ score for any sequence and any potential.
These average quantities are computed only once which saves significant
 amount of
computer time.

Energy of an individual conformation for pairwise Hamiltonian, 
is given by \ref{eq:ENER}. Hence average
energy for the set of conformations is:

\begin{equation}
\langle E(\mbox{\boldmath $\xi$}, \mbox{\boldmath $\Delta^{(k)}$} ) \rangle _k = 
\sum_{1\leq i < j\leq N} U(\xi_i,\xi_j) \langle \Delta_{ij} \rangle _k
\label{eq:AVR_ENERG}
\end{equation}

where $\langle \Delta_{ij} \rangle$ is average density of contacts between
residues number $i$ and $j$ in the set of alternative conformations.
\begin{equation}
\langle \Delta_{ij} \rangle _k = \frac{1}{K} \sum_{k=1}^{K} \Delta_{ij}^{(k)}
\end{equation}

Note that one can compute the matrix of average density of a contact
$\langle \Delta_{ij} \rangle$ only {\em once} for set of conformations and
then use this matrix to compute average energy for a sequence \mbox{ \boldmath $\xi$} and any potential {\bf U} (see eq \ref{eq:AVR_ENERG}).

Similarly for $ \sigma(E) $, 

\begin{equation}
\sigma^2 (E(\mbox{\boldmath $\xi$}, \mbox{\boldmath $\Delta^{(k)}$} )) _k = 
\langle E^2 \rangle _k - \langle E \rangle _k ^2  =
\sum_{1\leq i < j\leq N} \sum_{1\leq l < m \leq N} U(\xi_i,\xi_j) U(\xi_l,\xi_m) T_{ij,lm}
\end{equation}
where $T_{i,j,l,m}$ is contact correlator
\begin{equation}
T_{ij,lm}= \langle \Delta_{ij}^{(k)} \Delta_{lm}^{(k)} \rangle _k
- \langle \Delta_{ij}^{(k)} \rangle _k \langle \Delta_{lm}^{(k)} \rangle _k
\end{equation}
which depends only on the set conformations and can be computed in
advance for a given set. Once $\langle \Delta_{ij} \rangle$ and
$T_{ij,lm}$ are computed one can easily compute a value of $Z$ score
for a given sequence \mbox{\boldmath $\xi$}, conformation
\mbox{\boldmath $\Delta ^N$} and potential {\bf U}.  

\begin{equation}
Z(\mbox{\boldmath $\xi$}, \mbox{\boldmath $\Delta ^N$}, \mbox{\boldmath $U$}) =
\frac{ {\displaystyle \sum_{1\leq i < j\leq N} }
 U(\xi_i,\xi_j) ( \Delta_{ij} ^N - \langle \Delta_{ij}  \rangle _k ) 
}{
\sqrt{
{\displaystyle \sum_{1\leq i < j\leq N} } {\displaystyle \sum_{1\leq l < m \leq N}}
 U(\xi_i,\xi_j) U(\xi_l,\xi_m) T_{ij,lm}}
}
\label{eq:Zmain}
\end{equation}

\subsubsection*{Alternative conformations}
For each protein in the dataset we build an ensemble of alternative
conformations which contains conformations of the same compactness as
the native one, i.e. the same number of residue-residue contacts. In
fact, instead of generating a huge number of conformations, we
assume that (i) contacts in the alternative conformations are
distributed independently and uniformly and (ii) the number of
contacts is the same as in the native conformation. These assumptions
allow one to compute average density of contacts $ \langle \Delta_{ij}
\rangle $ and correlator $T_{ij,lm}$ as

\begin{equation}
\langle \Delta_{ij} \rangle = \frac{n}{n_{total}}
\end{equation}
and
\begin{equation}
T_{ij,lm} = \left\{ \begin{array}{ll}
\frac{1}{n_{total}^2} & \mbox{if $ij \neq lm$} \\
\frac{1}{n_{total}}-\frac{1}{n_{total}^2} & \mbox{if $ ij = lm $ }
\end{array}
\right.
\end{equation}
where $n$ is the number of contacts in the native conformation,
$ n_{total}$ is the total number of topologically possible contacts.

Then value of $Z$ score can be computed for each protein in the
dataset and a given potential {\bf U} using eq \ref{eq:Zmain}. Lattice model 
simulations show that sequences having 
low values of $Z$ are able to fold fast to
their native conformations (Abkevich {\em et al}, 1994   
Gutin {\em et al}, 1995)

To test our method of derivation and compare it with other techniques
we first turn to a simple lattice model which allows to test a potential
by performing {\it ab initio} folding of a protein starting from
random conformation.  Then we apply our method to derive the parameters
from the dataset of
well-resolved protein structures.

\subsubsection*{Sequence Design}
The aim of sequence design is to find a sequence which (for a given
potential) delivers low $Z$ score to a given conformation. The procedure
starts from random sequence with given amino acid composition. At each
step we choose two residues at random and attempt to permute
them. Change of $Z$ score ($\delta Z$) associated with this
permutation is computed. If this permutation decreases the value of
$Z$ score ($\delta Z <0$ ), then this permutation is accepted,
otherwise ($\delta Z >0$) permutation is rejected with probability $1-
\exp ( - \frac{ \delta Z}{T_{sel}})$. The procedure stops when either
no changes in sequence has occured in the last $1000 \dot N$ steps
or if a preset value of $Z$ score is reached ($Z_{target}$).
Using this procedure and setting different values $Z_{target}$ we are
able to generate sequences which provide the required value of $Z$ score
to a given conformation.


\section*{RESULTS}

\subsubsection*{Lattice model}
We consider a conformation of a protein chain as a self-avoiding
walk on a cubic lattice. Two amino acids which are not nearest
neighbors in sequence and located in the next vertices of the
lattice are said to be in contact. Energy of a conformation is
given by the equation (\ref{eq:ENER}).

\subsubsection*{Dataset of stable and folding proteins} 
The dataset of lattice proteins consists of 200 randomly chosen compact
conformations of 27-mer on 3x3x3 cube (Shakhnovich and Gutin, 1990;
Shakhnovich {\em et al}, 1991; Sali {\em et al}, 1994b; Socci and Onuchic, 1994). We derive the potential using
first 100 of the lattice proteins and test the derived potential for
the remaining 100  lattice proteins from the dataset.  We use potential
obtained by Miyazawa and Jernigan (MJ) as the true one. Using the true
potential for each native conformation in the dataset we design a
sequence which minimizes $Z$ score for this 
conformation (see (Shakhnovich and Gutin, 1993a,b) for
detailed description of design procedure).  
The stability and 
folding of each designed sequence is tested by Monte-Carlo
folding simulation, each starting from random coil 
(Shakhnovich {\em et al}, 1991; Sali {\em et al}, 1994) and reaching
their respective native conformations. 

\subsubsection*{Derivation of potential} 
To obtain the potential which minimizes $Z$ scores for model proteins, we
use Monte Carlo procedure in space of potentials (see above). Starting
from different random potentials Monte Carlo search converges fast, and
the resulting potential does not depend on starting random
potential. The procedure converges to a unique potential
even at zero optimization temperature $T_{opt}=0$.  This shows that
there is only one minimum in  space of potentials in our model. 
This guarantees that the derived 
potential provides the global minimum to the target function $ \langle
Z \rangle _{harm}$.

\subsubsection*{Derived vs true potential} 
\begin{figure}
\epsfxsize=6in
\epsffile{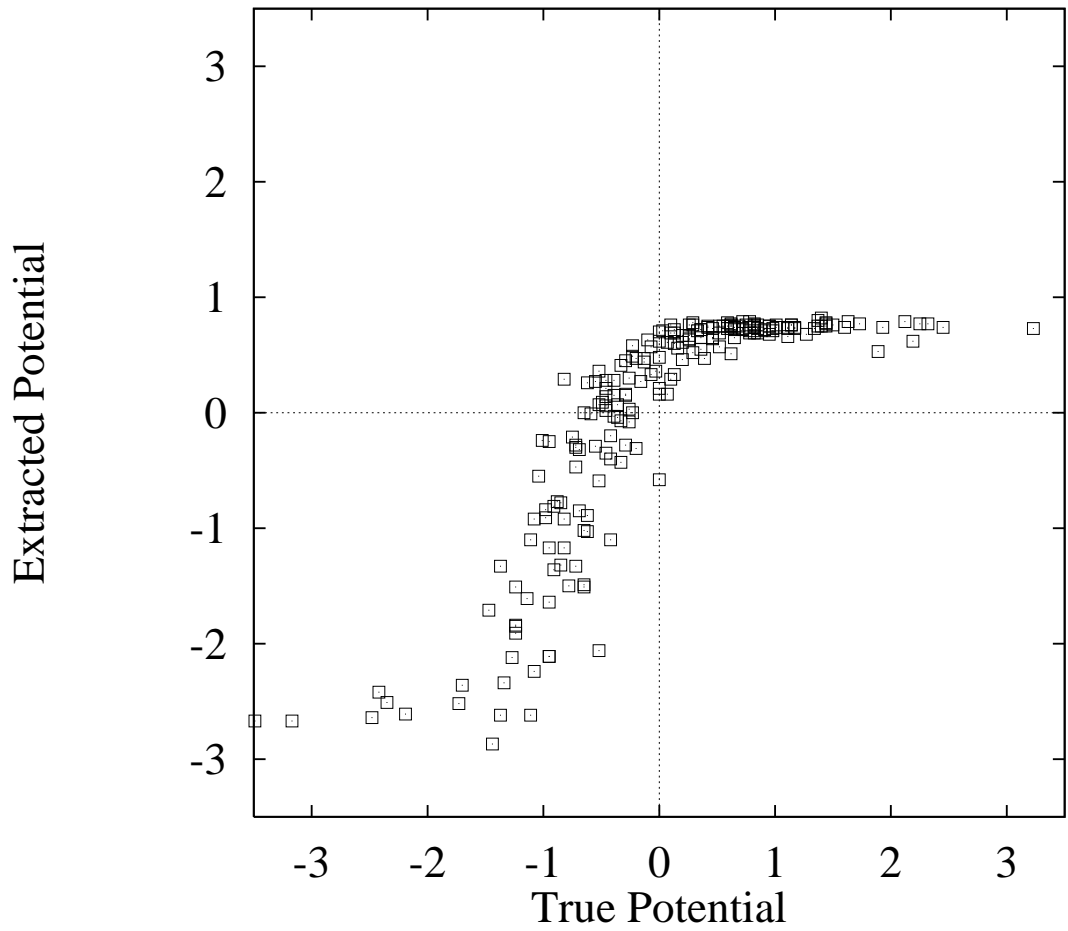}
\vspace{0.5in}
\caption{Derived potential vs ``true'' potential for the 
lattice model}
\label{fig:derived_vs_true.lattice}
\end{figure}

The potential obtained this way has been 
compared with the real one in
several ways.  Figure \ref{fig:derived_vs_true.lattice} presents 210
values of interactions in the derived potential 
vs the same values for the
true potential. Correlation $r=0.84$ shows that our
method is able to find the true potential. \marginpar{Fig.1}

The values of energy for attractive interactions $(U(\xi, \eta) <0)$ are
predicted much better than those for the repulsive ones $(U(\xi, \eta)
>0)$. Attractive interactions stabilize the native conformations and
appear much more frequently among native contacts. Repulsive
interactions, in contrast, are very rear among native contacts and
therefore the statistics is much poorer for them. Some 
repulsive contacts  cannot be found 
in the dataset of model proteins. In contrast, 
contacts between {\em all}
amino acids are present among native contacts 
in real proteins (see below). The absence of contacts
between some types of amino acids in the model dataset is the result
of very strong sequence design. The design finds a sequence which
provides very high stability of the native conformation in the given model,
and by doing
so it eliminates repulsive contacts which destabilize the native
conformation. This observation is the first indicator that native
sequences do not appear as well designed for stability {\em in terms
of our model} (contact pairwise potential, $4.5 A$ cutoff of
residue-residue interactions etc). 

\subsubsection*{Ab initio folding with derived potentials}
Ability of the derived potential to fold model proteins was
tested by their {\it ab initio} folding.  

Folding simulations were carried out using standard Monte-Carlo
method for polymers on a cubic lattice.  The detailed discussion
of lattice Monte-Carlo simulation technique, its advantages and
caveats is given in many publications (see e.g.
Sali {\em et al}, 1994, Socci and Onuchic, 1994). 
Each simulation started from a random
coil conformation, proceeded at constant temperature and lasted
about 5 times longer than the mean folding time

\begin{figure}
\epsfxsize=6in
\epsffile{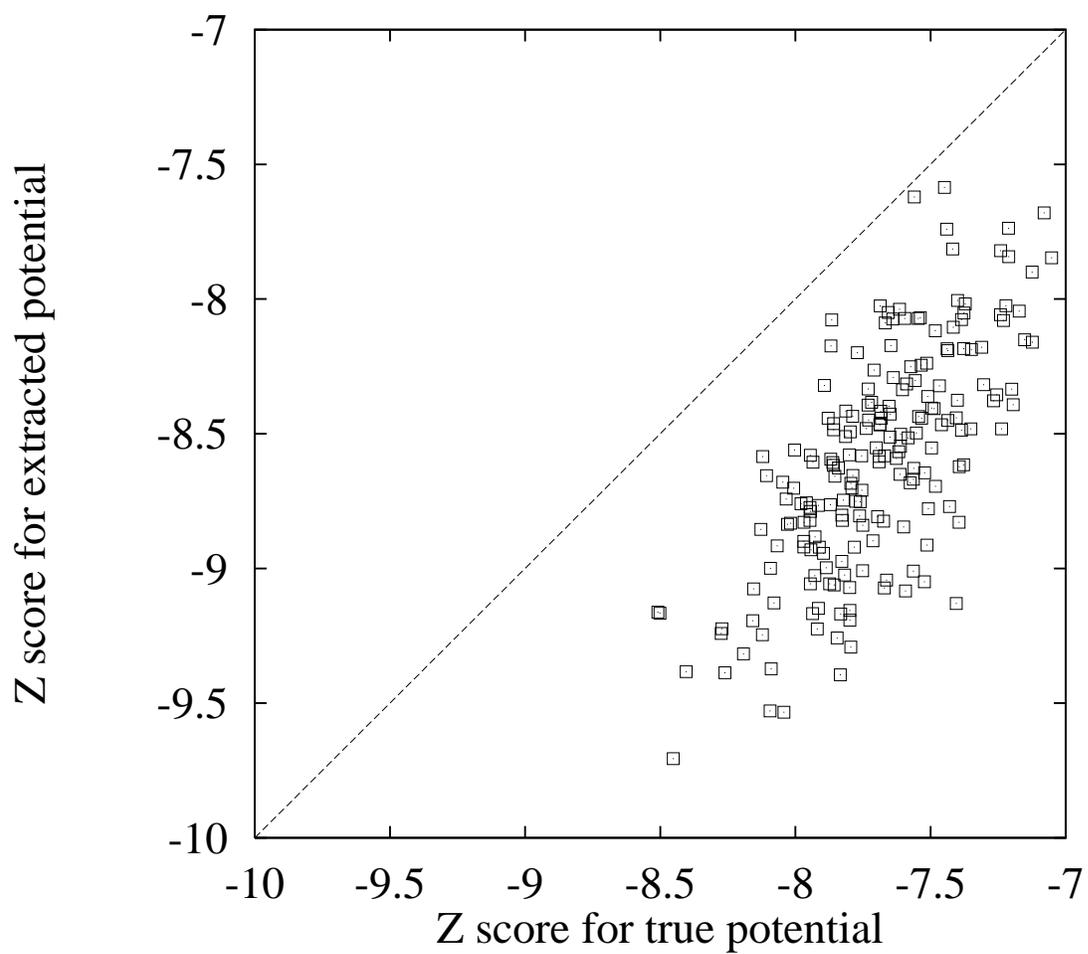}
\vspace{0.5in}
\caption{$Z$ score of model proteins with the 
derived potential vs $Z$ score 
with the``true'' potential.}
\label{fig:z_vs_z}
\end{figure}

All tests were
performed for proteins which were not used for the derivation
procedure. First we compare $Z({\text {\bf U}}_{der})$ values provided
by derived potential with $Z({\text {\bf U}}_{true})$ values for the
true potential. Figure \ref{fig:z_vs_z} presents $Z({\text {\bf
U}}_{der})$ as a function of $Z({\text {\bf U}}_{true})$. Derived
potential provides almost the same or even lower values of $Z$ score
for all proteins. \marginpar{Fig.2}

\begin{figure}
\epsfxsize=6in
\epsffile{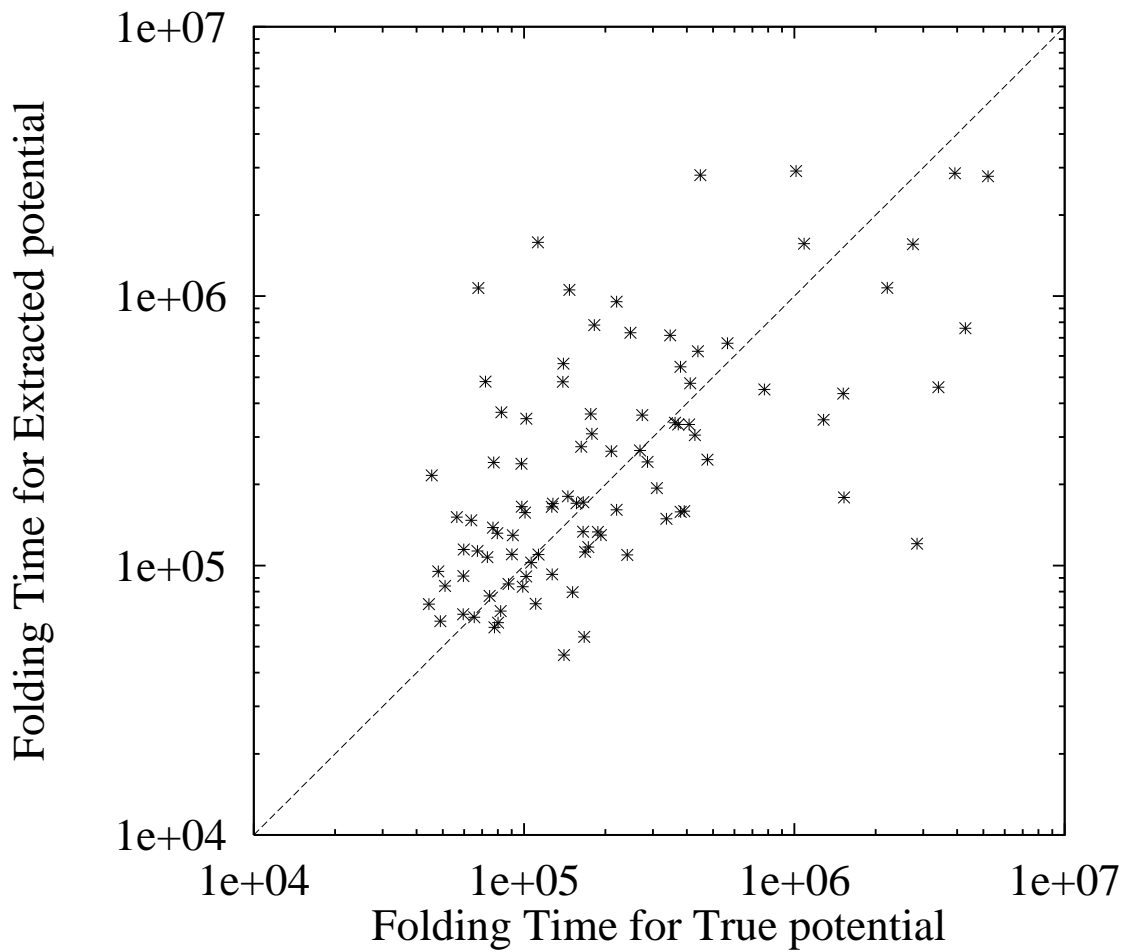}
\vspace{0.5in}
\caption{Folding time with derived potential vs folding time with the``true'' potential for the lattice model. 100 lattice model proteins, not used
for derivation of potentials were taken for Monte-Carlo folding simulations.
Folding ``time'' is measured in Monte-Carlo steps required to reach the
native conformation.}
\label{fig:mfpt}
\end{figure}

We also define folding time for each protein as mean first passage
time, when the native conformation was reached first. Time
is measured in MC steps. 40 runs were 
performed for each protein for
both ``true'' and derived potentials.  Simulations were run at 
temperature $T=0.7$.  Figure \ref{fig:mfpt}.  presents 
the scatter plot of folding time
obtained for the derived potential vs folding time for the ``true''
one. All proteins which fold with the ``true'' potential fold also with the
derived potential exhibiting approximately the same folding time. \marginpar{Fig.3}

The folding test proves that the derived potential is able to provide fast folding for proteins with
well-designed sequences.

\subsubsection*{Effect of the database size} 

How sensitive is the derived potential to the number of proteins used for
the derivation? How many proteins are required to obtain the potential
similar to the true one. To address these questions we performed
derivation of potential for databases containing different number of
proteins.

\begin{figure}
\epsfxsize=5in
\epsffile{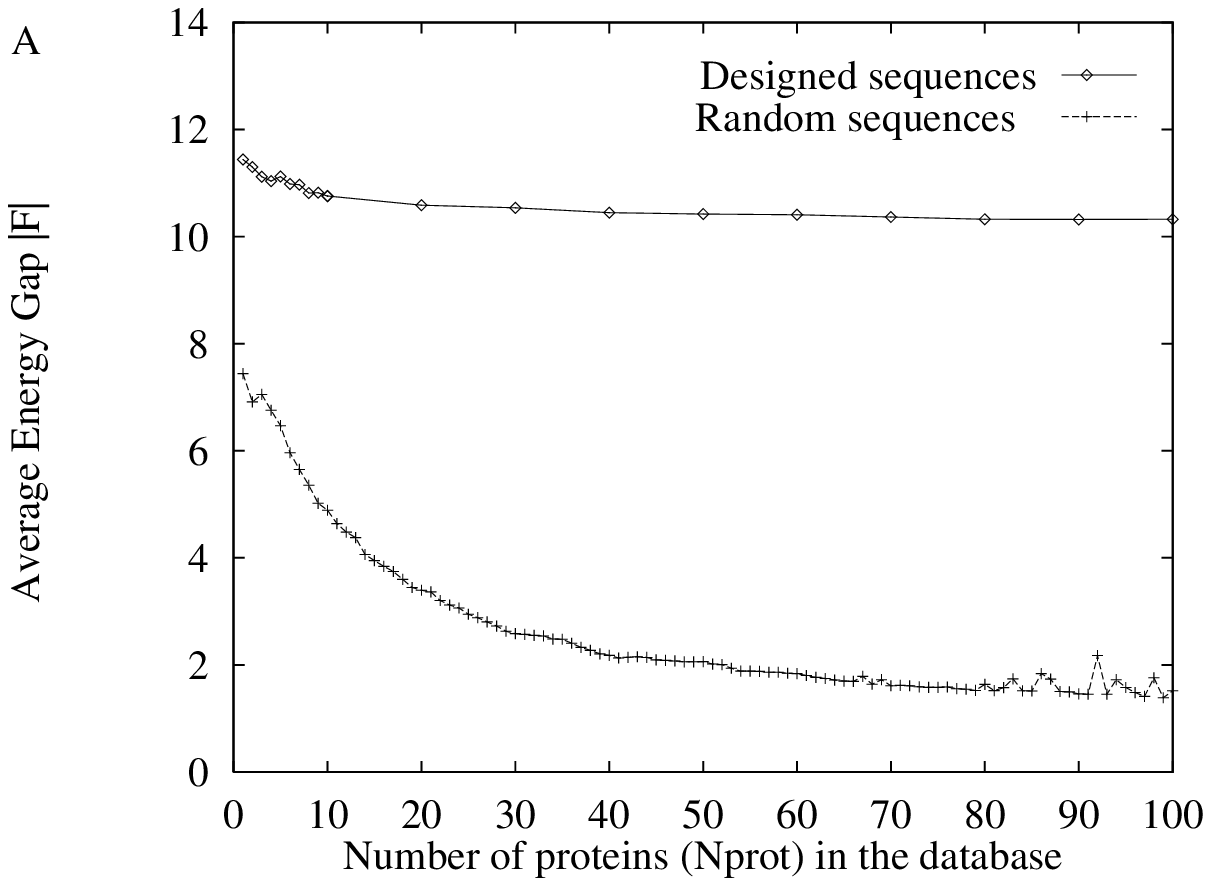}
\epsfxsize=5in
\epsffile{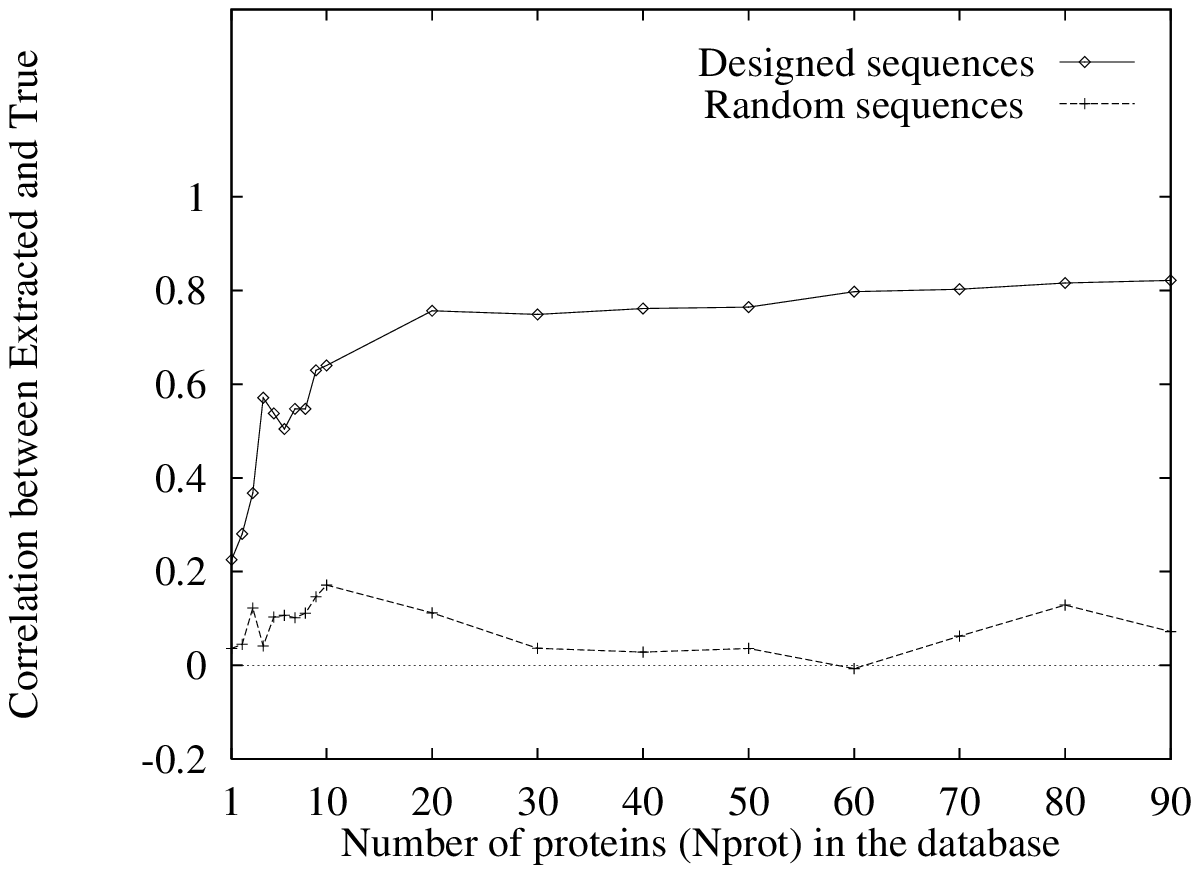}
\vspace{0.5in}
\caption{Effect of the database size, used for derivation of the potential, 
on average energy gap (a) and convergence test (b), for lattice model proteins}
\label{fig:converge}
\end{figure}

For the database containing $N_{prot}$
proteins we derive a potential using the technique described above and
compute the average score (energy gap) $ |F| = | \langle Z \rangle _{harm} | $
provided by the derived potential. (Fig.\ref{fig:converge}a). 
We also compute correlation of the true potential
and the one obtained for $N_{prot}$ proteins (see Figure
\ref{fig:converge}b). \marginpar{Fig.4}

For few ( $1..5$ ) proteins one can obtain a potential which
provides very large energy gap for these proteins. This potential,
however, fails to provide reasonable gap for other proteins and is not
 similar ($r=0.2..0.5$) to the true potential.  As the number
of proteins in the database increases, the average energy gap decreases
approaching rather high constant value of ($|F|=1.6$). Correlation
between derived and true potentials approaches constant value
$r=0.85$. To ensure that derived 
potential converges to a meaningful value 
as the number of
proteins in the database increases, 
we compare potential obtained for 
$N_{prot}$ proteins with the one 
obtained for all $100$ proteins. Correlation
between these potentials as a function of $N_{prot}$ is shown on
Figure \ref{fig:converge2}.  Clearly, as the 
number of proteins in the
database increases, correlation between derived potentials approaches
$1$ and, hence, potential converges to a unique solution.

The results of this procedure clearly demonstrate the stability of our
procedure. It is also important that the potential which is 
highly correlated
with the true one ($r=0.8$) can be obtained with only $40..50$
proteins, which is of the order of the size of the database of
non-homologous stable disulfide-free proteins available from the PDB.

\subsubsection*{Is there enough parameters? 
\\Is there too many parameters?}  An important issue 
 is whether the number of parameters adjusted in the potential
is sufficient to provide large gap for 
all proteins with designed sequences.
E.g. two-letter (HP) models
are too nonspecific to make native structure
unique: for any sequence 
many conformations of 3-dimensional HP heteropolymers
have the same lowest, global minimum energy.
Native state in such models is
not unique in most cases; correspondingly  no sequence,
random or designed
can have any energy gap 
in HP models (Yue {\em et al}, 1995).

On the other hand the number of parameters should not be too 
large because in this case, it will always be
possible to find a ``potential'' 
for which all members of the database used in the derivation
 have low energies, but the resulting potential has
nothing to do with the ``true'' potential and it will not provide low
energy to proteins which are not members of the derivation database.

 More specifically, the question is whether the problem of finding
parameters is under-determined or over-determined, i.e how the number of
independent functions to minimize $Z$-scores of individual proteins 
is related to the number of
independent parameters. For an over-determined problem the number of
functions/constraints is greater than the 
number of parameters and, hence,
there is no solution which minimizes well all the functions. This is
not the case for our designed sequences, since there is ``true'' potential
which provides large enough gap for all proteins. Below we address
this question for native sequences of real proteins.  
If the problem is under-determined,
then the number of functions/constraints 
is less than the number of
parameters and one can find infinitely many solutions minimizing all
functions. This is the case when the number of proteins in the database is
small.  As we have shown above, the 
potential derived for few proteins
provides for them the average energy gap greater 
than provided by the ``true''
potential, but it shares no similarity 
with the ``true'' potential.

However, as the number of proteins in the database increases, average gap
approaches that for the true potential and the derived potential  becomes
very similar to the true one. 
To ensure that we {\em do not have too many parameters} we made the control
procedure with randomly shuffled sequences.

\subsubsection*{Randomly shuffled  sequences: an essential control} 

As a control we carried out the derivation procedure for our database of
model proteins using  shuffled sequences instead of the designed ones
for each protein. In this case one should not expect that there
exists any potential which makes all the native structures
to be of low energy for randomly shuffled sequences, i.e in this case
our procedure should not lead to any meaningful solution.
What happens in this case?

\begin{figure}
\epsfxsize=6in
\epsfysize=5in
\epsffile{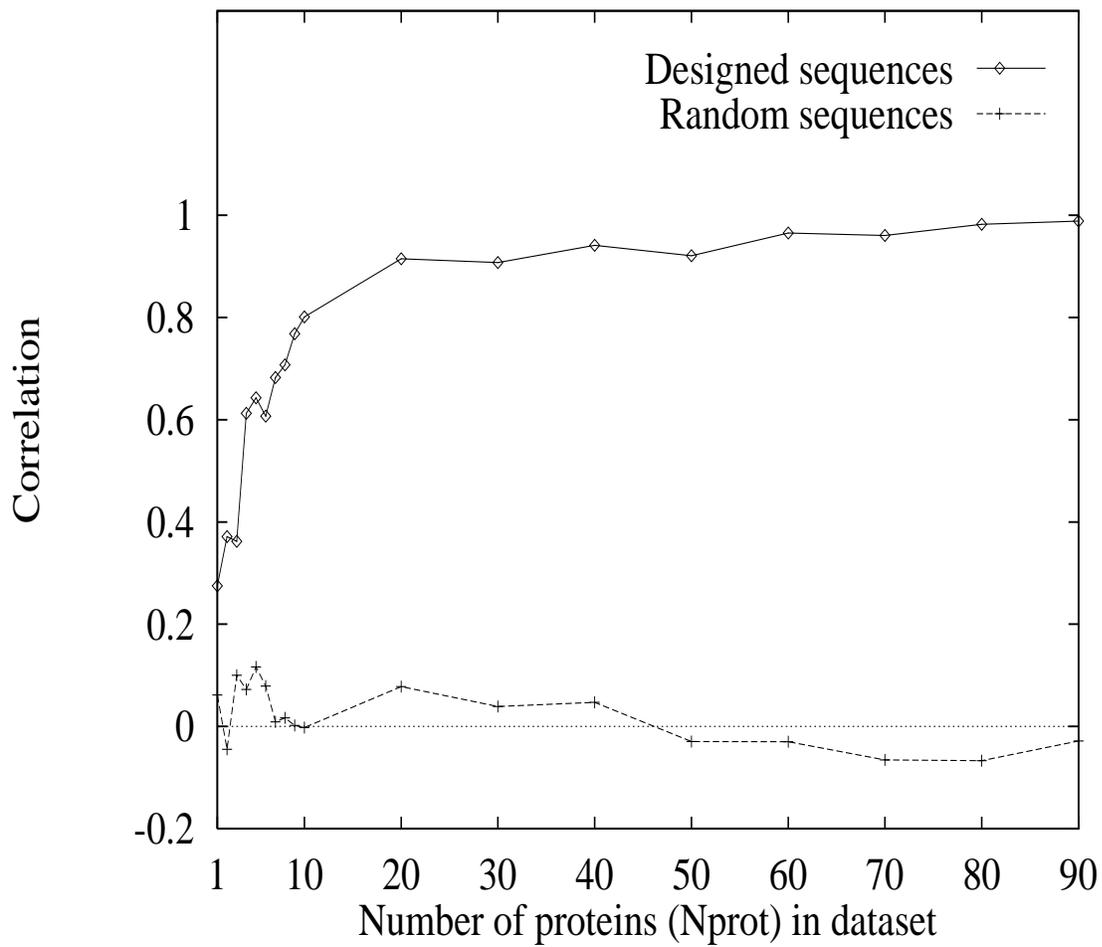}
\vspace{0.5in}
\caption{Convergence of potential for lattice model proteins. 
Correlation between potentials
derived from all  
$100$ model proteins and potential derived from $ N_{prot} <100 $
proteins is shown as a function of $N_{prot}$}
\label{fig:converge2}
\end{figure}

Again, for a a few ( $1..5$ ) proteins one can find
a potential which provides large enough energy gap ($|F|=0.8..1.2$)
for randomly shuffled sequences (see Figure
\ref{fig:converge}a). However, in contrast to the designed sequences,
average energy gap drops substantially to a marginal level of
$|F|=0.2$ as number of proteins in the database increases. Clearly,
there is no correlation between the true potential and potential
derived for database with shuffled sequences (see Figure
\ref{fig:converge2}). There is also no correlation ($r=0.0$) between
potential obtained for $100$ proteins with shuffled sequences and
potentials obtained for $N_{prot}<100$ of these proteins. Hence, the
procedure does not converge to any potential for proteins with
randomly shuffled sequences.

Consequently, no pairwise potential can provide stability simultaneously 
to all of native conformations with the shuffled sequences.

Comparison of the results  for designed  sequences 
with the control case of shuffled sequences suggests 
that the problem of finding a pairwise potential
is not under-determined, i.e. $210$ parameters of the potential 
are sufficient 
to provide large gap for designed sequences and are not sufficient
to provide large gap for any 
pair of sequence and conformation. \marginpar{Fig.5}

Note that the procedure is able to distinguish between designed and
randomly shuffled sequences without prior information about
the potential used for the design.Designed sequences show the convergence of
average energy gap $|F|$ to a level of $F=1.4$ as number of proteins
increases. Potential is converging which is seen from high correlation
between potentials obtained for different number of proteins
in the dataset. In
contrast, no convergence to a single potential is observed for
proteins with randomly shuffled sequences (see Figure \ref{fig:converge2}). The target function $F$ approaches small values of $0.2$ as number of
proteins in the database increases.Where are the native proteins on this
scale? Do they behave more like designed 
or like randomly shuffled sequences?

\subsection*{Native proteins}

\subsubsection*{The model} 

We built a database of proteins with less than 25\% of sequence
homology, longer than 50 and shorter than 200 amino acids.  The
database contains 104 proteins, listed below (in pdb-code names): 1hcr
1cad
1enh
1aap
1ovo
1fxd
1cse
1r69
1plf
2sn3
1bov
1mjc
1hst
1hyp
1ubq
4icb
1pk4
1poh
1aba
1lmb
1cyo
1brs
1fna
1mol
1stf
1gmp
1frd
1hsb
1ida
1plc
1aya
1onc
1sha
1fus
1psp
1fdd
256b
1acx
1bet
1fkb
1pal
2sic
1brn
2trx
1ccr
2msb
1dyn
1c2r
1etb
1gmf
2rsl
1paz
1rpg
1acf
2ccy
3chy
135l
1aiz
1rcb
1adl
1bbh
1slc
1eco
2end
4fxn
1ith
1cdl
1flp
2asr
1ilr
1lpe
1hbi
1bab
1lba
1mba
8atc
1ash
2fx2
2hbg
2mta
1f3g
1ndc
1aak
1cob
4i1b
1mbd
2rn2
1esl
1hfc
1hlb
1pnt
1hjr
4dfr
119l
3dfr
2cpl
5p21
1rcf
9wga
2alp
1fha
1bbp
2gcr
1hbq. 
We use this database to derive
the potential which maximizes average energy gap $ | \langle Z \rangle
_{harm} | $. 
We define a contact between two aminoacids when
the distance between their nearest heavy atoms is less than cutoff
value 4.5A.  
In contrast to the
lattice model, real proteins have different length and different
number of contacts.  These factors affect the value of $Z$ score. To
account for the increase of $Z$ with protein length we introduce the
following normalization:

$$ Z_{norm}= \frac{Z}{\sqrt{n_{nat}}} $$

where $n_{nat}$ is number of native contacts. Normalized values
$Z_{norm}$ are used to compute $ F_{norm} = \langle Z_{norm} \rangle
_{harm}$ harmonic mean. Our criterion 
$\langle Z_{norm} \rangle_{harm}$  overemphasizes poor scores;
therefore it is very 
sensitive to proteins in the database which are 
more random-like, and their presence in the dataset
can distort the resulting potential.
To avoid this difficulty we selected proteins for the dataset, which
we believe are stabilized by similar physical forces (hydrophobic,
electrostatic, H-bonds, etc) and avoided 
proteins which are stabilized by other factors 
such as disulfides or coordinated metals, heme groups etc. \marginpar{Fig.6}

\begin{figure}
\epsfxsize=6in
\epsfysize=4in
\epsffile{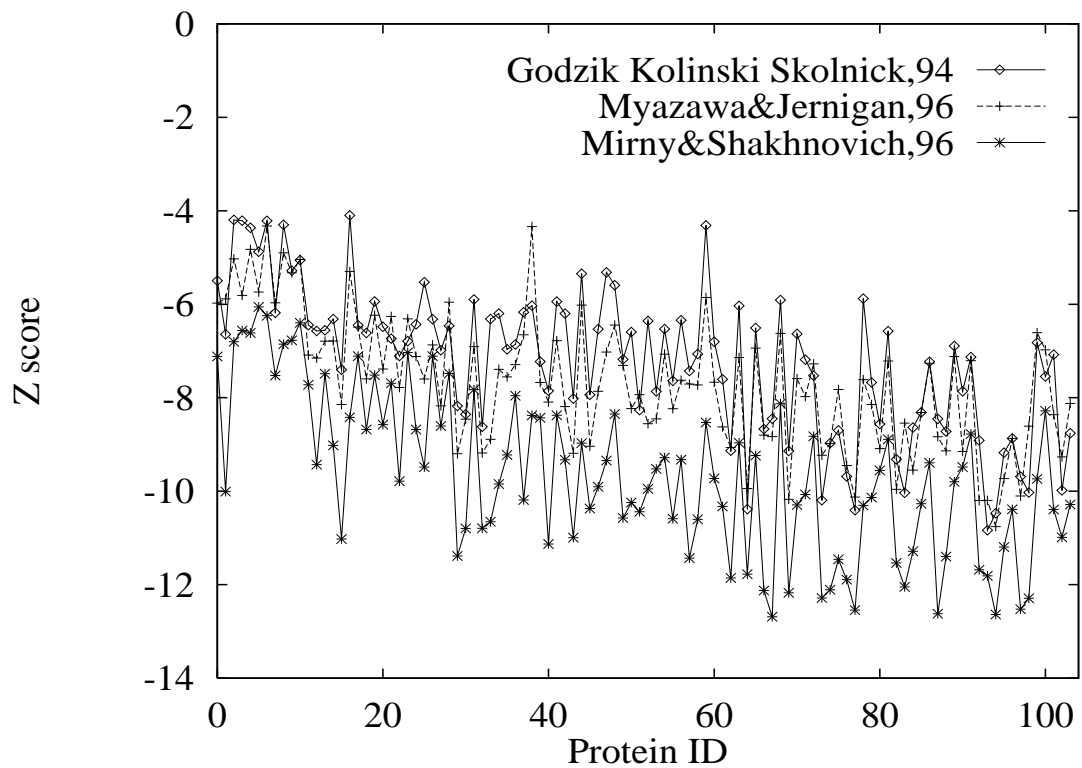}
\vspace{0.5in}
\caption{$Z$ score of native proteins with different potentials.
Proteins are arranged in order of increasing length; this explains the systematic trend of decrease of $Z$-score as protein ID \# increases}
\label{fig:z_vs_id}
\end{figure}

Figure \ref{fig:z_vs_id} presents values of $Z$ score obtained for the
derived potential. Although our method finds the potential which maximizes
average energy gap ( $ |Z| $ ) simultaneously for all proteins, the
values of the gap obtained for real proteins are rather small. This
indicates that in the framework of the model we use (contact pairwise
potential, $4.5 A$ cutoff of residue-residue interactions etc) no
pairwise potential can provide high stability simultaneously to all
native proteins.

\subsubsection*{How good is the model for native proteins?} 

The potential derived from native proteins converges 
to  certain value of Z-scores. Is this value large or small?
To answer this question we should compare it with two
limiting cases: a) when the functional form of the 
energy function is ``exact'', and sequences are well-designed
for this energy function and b) with randomly shuffled sequences.

For each protein in the dataset of native proteins we
design a sequence using MJ potential as  ``true'' potential and
preserving amino acid composition of the native sequence.  Then we
derive a potential for a subset containing $N_{prot}$ proteins from
the database. The derivation is performed for the proteins built of
$(i)$ the native structures with their native sequences, $(ii)$ the
native structure with sequences designed for them; $(iii)$ and the
native structures with randomly shuffled sequences.

\begin{figure}
\epsfxsize=6in
\epsfysize=4in
\epsffile{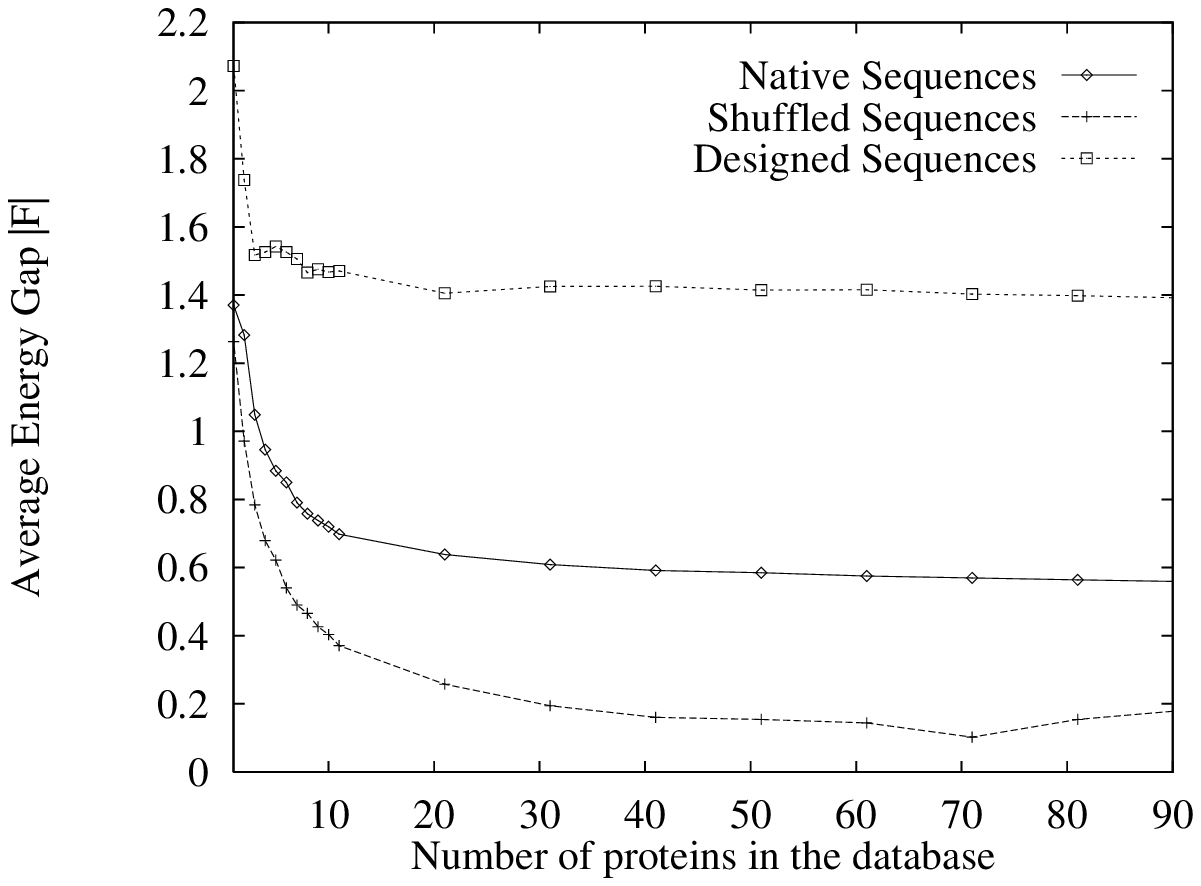}
\vspace{0.2in}
\epsfxsize=6in
\epsfysize=4in
\epsffile{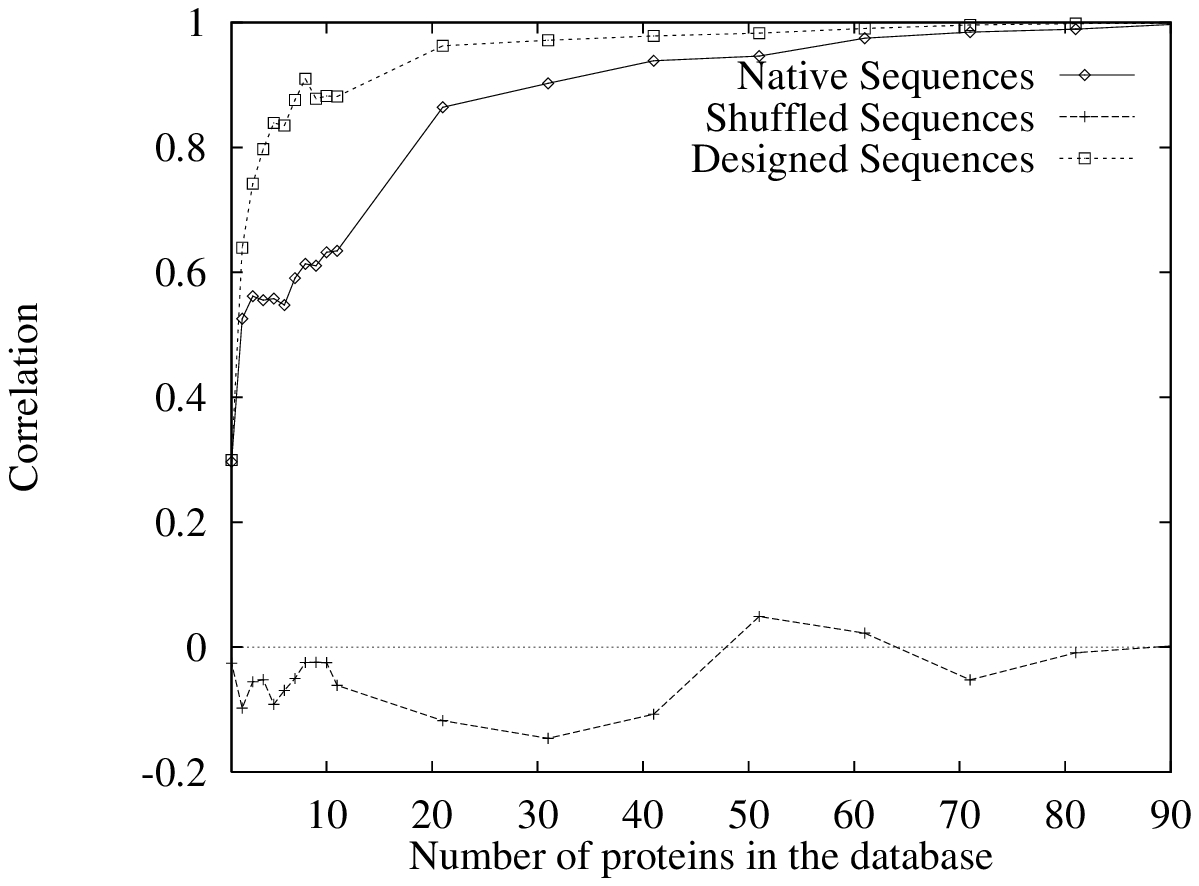}
\vspace{0.1in}
\caption{Native proteins. (a) Effect of the 
database size on the energy gap  and
(b) Converegency test: Correlation between the potential derived using smaller database and the potential
derived using all 104 proteins in the database}
\label{fig:convergepdb}
\end{figure}

Figure \ref{fig:convergepdb} presents average normalized energy gap
value $|F_{norm}|$ obtained for all three sets of sequences as a
function of number of proteins in the database. Similar to the lattice
model (see Figure \ref{fig:converge}a), designed sequences reach
high values of the gap $|F|=1.2$ whereas for random sequences
$|F|=0.2$. Derivation of potential for native sequences yields
$|F|=0.56$ which is considerably less than the gap provided for the same
structures with designed sequences. \marginpar{Fig.7}

Another important property of designed sequences is that 
the derived potential converges to a single
potential as number of proteins in the database increases. Randomly
shuffled sequences, in contrast, lack this
convergence.  The criterion of this convergence is correlation between
potentials derived using $100$ proteins and potential derived using
$N_{prot} < 100$ proteins.  Correlation between potentials obtained
for the database of native proteins as a function of number of
proteins in the database is shown on Figure \ref{fig:convergepdb}b.

In contrast to randomly shuffled sequences, native sequences as well
as designed sequences provide convergence to a single potential as number
of proteins used for the derivation increases. Hence, we are able to
find a potential which maximizes, for the given model, the  
energy gap for all native
proteins simultaneously. This result clearly demonstrates that the
model energy function used in this study of proteins is meaningful
and reflects some essential interactions, but not all, since there is a
pronounced difference
between F-values for designed and random sequences.

Since our method of derivation maximizes $|Z|$ scores for all proteins,
there is no potential which can provide greater $|Z|$ score, for the
studied Hamiltonian (pairwise interaction potential) than this
method does. Our results demonstrate that very moderate $|Z|$ scores
can be obtained using pairwise potential for native proteins and no
potential can increase values of $|Z|$ for them. However, other models
utilizing different protein structure representation or different form
of potential can be more efficient in providing large energy gap for
native proteins. Using our procedure one can compare different
models quantitatively and select better one which provides larger
energy gaps for native proteins.

\subsubsection*{Can we derive a potential from the dataset of 
poorly designed sequences?}
Using the lattice model and native proteins, 
we demonstrated that our procedure is able to reconstruct ``true''
potential sufficiently accurate if sequences in the database 
are well designed. Is this requirement too restrictive? 
How well can we
reconstruct potential for proteins with poorly 
designed sequences?

To mimic poor design of native proteins observed in our model, for
each protein structure, we designed sequences which provide the same
value of energy gap as the native sequence does. Design is performed
using MJ potential as the ``true'' one. Then we derived 
the  potential for
these poorly designed sequences and 
compared them with the ``true'' potential.

\begin{figure}
\epsfxsize=7in
\epsffile{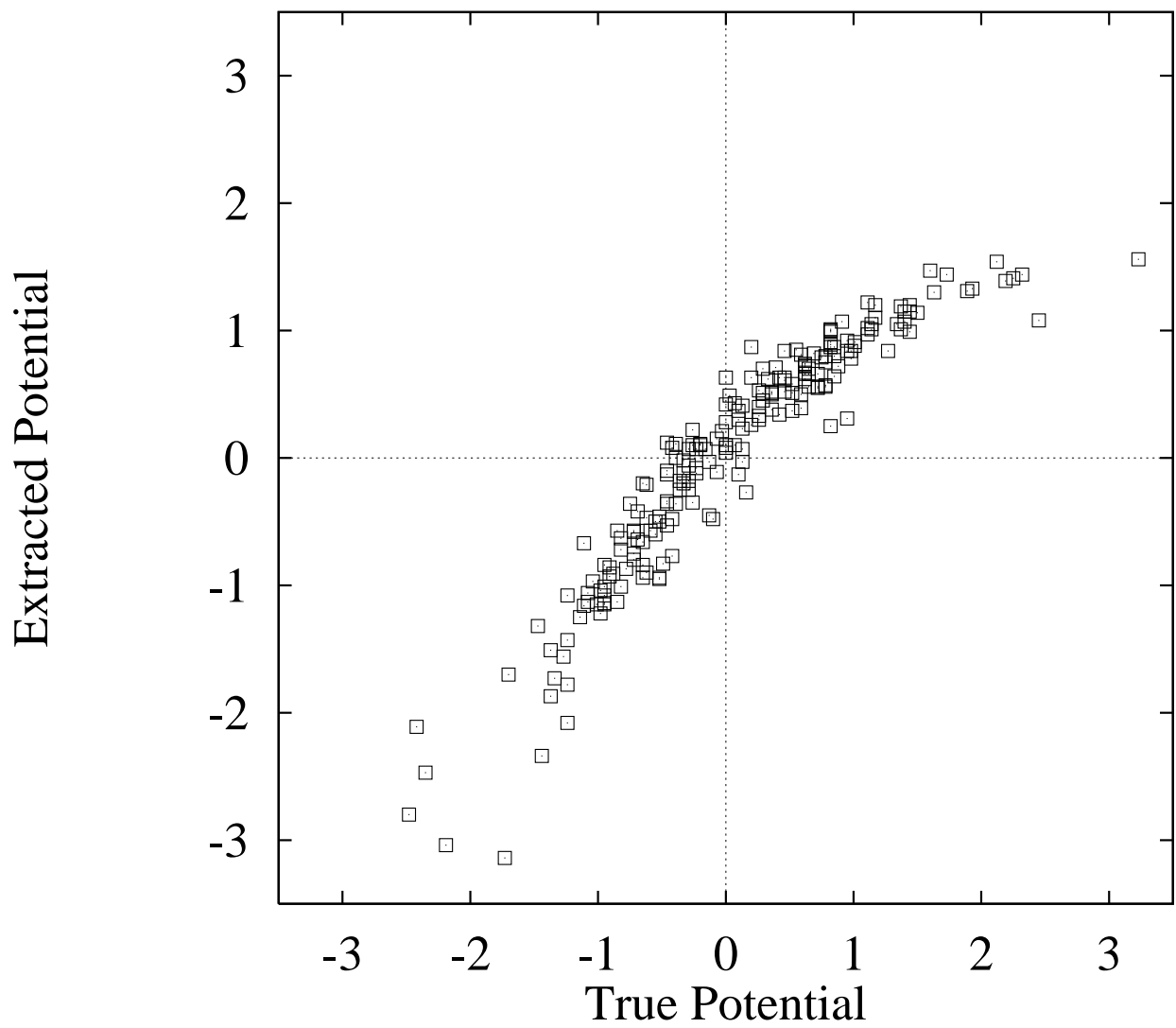} 
\vspace{0.5in}
\caption{Derived potential vs ``true'' potential for the native proteins}
\label{fig:derived_vs_true.pdb}
\end{figure}

Figure \ref{fig:derived_vs_true.pdb} presents the scatter plot
of interaction
energies for the obtained potential plotted against the 
same values of the
``true'' potential. The result shows that our procedure reconstructs
potential for poorly designed sequences very well providing correlation
$r_{poor}=0.91$ with the true potential. Poor design of native
sequences in our model does not affect quality of reconstruction of
the true potential. \marginpar{Fig.8}

\subsubsection*{Can poorly-designed proteins fold?}
As we have demonstrated above, native proteins are rather poorly
designed in terms of the pairwise potential. The best possible
pairwise potential provides rather small energy gap to the native proteins,
which is characterized by the typical value of $Z_{norm}=0.5..0.7$.
Well designed proteins have, in contrast, $Z_{norm}=1.2..1.4$ and may be
able to fold to their native conformation, as lattice model simulations suggest. The question is whether
poorly designed sequences can fold as well.

To address this question we turn to the ``ideal'' lattice model and build
a dataset of $200$ poorly-designed proteins. Proteins in this dataset
are designed to have $Z_{norm}=0.5..0.7$ . Then we derived potential
for this dataset as described above and performed folding simulations
for all sequences using ``true'' and derived potential. The result is
that no protein was able to fold to its native conformation neither
with derived nor with the ``true'' potential.

In {\em all} cases there was a conformation which has an energy below
the energy of the native conformation. Hence,

\begin{itemize}

\item the native conformation is {\em not} the global energy minimum
for a poorly designed protein;

\item poorly designed proteins are unable to fold to their native
conformations in {\it ab initio} folding simulations.

\end{itemize}

\subsubsection*{Fold recognition of poorly-designed proteins.}

Sampling techniques which are more constrained to protein-like
conformations (Finkelstein and Reva, 1991; 
Jones {\em et al}, 1992; Wodak and Rooman, 1993)
 can, however, recognize the native and native-like
folds among small enough pool of alternative conformations. The 
success
of different pairwise potentials for the fold recognition shows that
this sampling technique works quite well even for poorly designed
proteins. Using our set of poorly designed sequences we made fold
recognition tests for all lattice model proteins by threading sequence of each
protein through 200 alternative conformations. Only 3 out of 200
sequences recognized non-native conformation as those of the lowest
energy. This result is in contrast to previous observation that for
any protein in this set native conformation was not the global energy
minimum. Hence, the only reason why fold recognition works for 197
proteins is that a set of decoys was not too large and 
representative so that the native
conformation had the lowest energy. 

Not surprisingly, the 
comparison of the results of {\it ab initio} folding simulations and fold
recognition indicates that folding is a much more complicated problem
than fold recognition since much greater energy gap is required for
success in folding than in fold recognition.  

The question whether poorly designed proteins can be used for
recognition of the native fold in threading experiments yet to be
studied systematically.

\subsection*{Comparison with other potentials and techniques for extraction of potential.} 

Several knowledge-based 
techniques for derivation of potentials from native protein
structures have been suggested. It is important to compare our
potential with other pairwise potentials,
and our method of derivation with other methods. Figure
\ref{fig:z_vs_id} presents $Z$ scores computed for proteins of our
database using our potential and two other potentials taken from the
literature. Clearly our potential provides significantly lower values
of $Z$ score for all proteins in the database. Two other potentials
perform good as well providing rather low $Z$ scores. Although
potentials were obtained using different techniques, the overall
profiles of $Z$ score for this dataset of proteins are very similar
for all three potentials, i.e. when a protein has low $Z$ score with
one potential it usually has low $Z$ score with another one. High
correlations between values of $Z$ score provided by these three
potentials for the same set of proteins ($r_{MJ,GKS}=0.83$
$r_{GKS,MS}=0.86$ and $r_{MJ,MS}=0.83$) indicate that high or low
value of $Z$ score is a property of a protein itself irrespective of
potential used. Different proteins are known to have different
stability, i.e.  different quality of design, which is displayed by
high or low value of $Z$ score.

\subsubsection*{Comparison with other techniques for extraction of potential.}

It is  important to compare not only potentials by themselves but also
the techniques for derivation of potential. Our 'ideal' lattice model
is very useful for this purpose. We apply different techniques to the
same set of lattice proteins and test obtained potentials in the same
way we did this for our technique.

Here we compare four techniques for derivation of potential. First two
are widely used statistical knowledge-based method to derive energy of
residue-residue and residue-solvent interactions. Knowledge-based
techniques are reproduced following 
Miyazawa and Jernigan (1985, 1996) (MJ) and
Hinds and Levitt (1995) (HL). The third 
tested technique is the procedure
suggested by Goldstein {\em at al} (1992) (GSW). This procedure is somewhat
similar to out method since the potential is obtained to maximize ratio
$\frac{T_c}{T_f}$, which is similar the $Z$ score we use. Goldstein 
{\em et al} found analytic expression for potential which maximizes
$\frac{T_c}{T_f}$ for one protein. To find the potential for a set of
proteins they used averaging which is not justified  but it
yielded good results. We followed the procedure described in 
(Goldstein {\em at al} 1992) to
test their  technique. Note that both GSW 
procedure and ours are optimization techniques,
whereas HL and MJ are statistical knowledge based 
ones. \marginpar{Fig.9}

\begin{figure}
\epsfxsize=7in
\epsfysize=4in
\epsffile{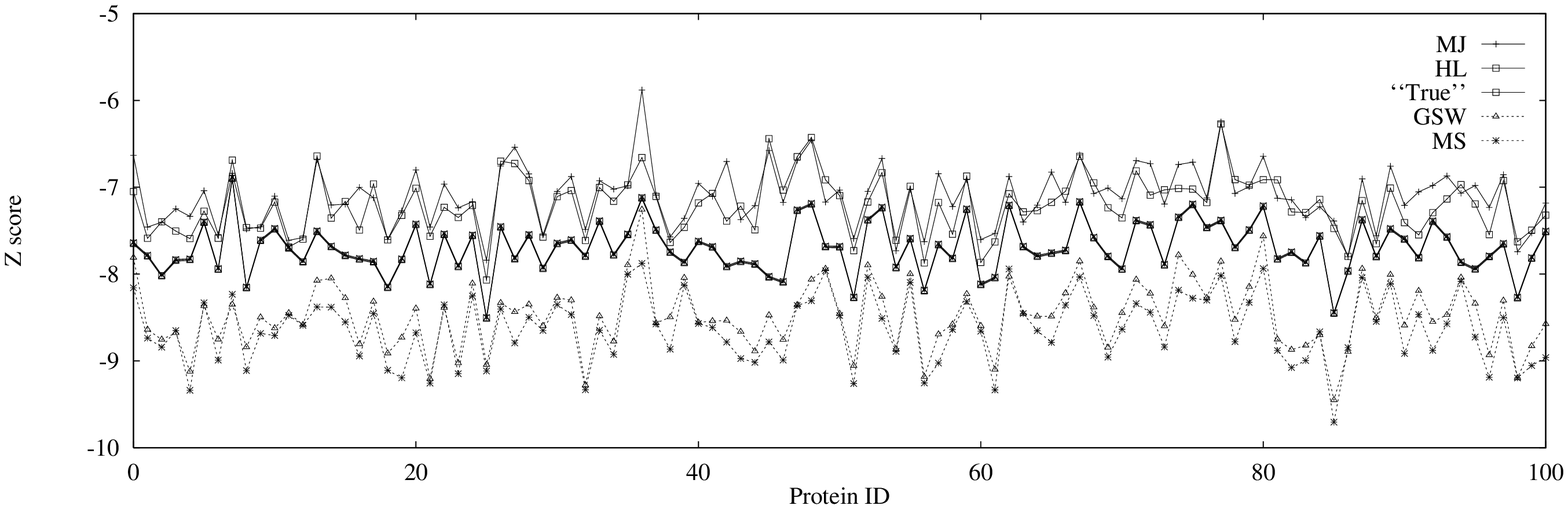}
\vspace{0.5in}
\caption{$Z$ score for 100 test lattice proteins  with potentials derived  by different techniques
from 100 ``database'' lattice proteins.
HL - Hinds \& Levitt, MJ - Miyazawa \& Jernigan, GSW - Goldstein {\em et al}, MS - this work}
\label{fig:z_vs_id2}
\end{figure}

The results for different techniques are summarized in 
Table \ref{tab:kb_wg_ms}.  
Our potential is aimed to minimize harmonic 
mean $Z$ score
and, as expected, provides the value of $\langle Z \rangle _{harm}$
lower than other potentials. GSW procedure gives only slightly higher
values of mean $Z$ which proves that both optimization techniques are
powerful enough to provide large energy gap for proteins of a dataset.
Knowledge-based techniques provide large energy gap as well.  The
drastic difference between knowledge-based technique and optimization
techniques becomes transparent when we compare $Z$ scores obtained for
different derived potentials with $Z$ scores provided by the ``true''
potential (see Fig. \ref{fig:z_vs_id2}). Both optimization techniques
provide $Z$ scores which are lower than $Z$ for the ``true''
potentials.  Knowledge-based techniques, in contrast, provide $Z$ scores
higher that those for the ``true'' potential.  Hence, knowledge-based
potentials provide smaller energy gap than the ``true'' potential does,
whereas potentials obtained by optimization deliver the energy gap
which is greater than those for the ``true'' potential. The decrease of energy
gap by knowledge-based potentials can be crucial for {\it ab initio} folding,
especially for weakly designed proteins which have rather small gap even with
the ``true'' potential. \marginpar{Table 1}

\begin{figure}
\epsfxsize=6in
\epsffile{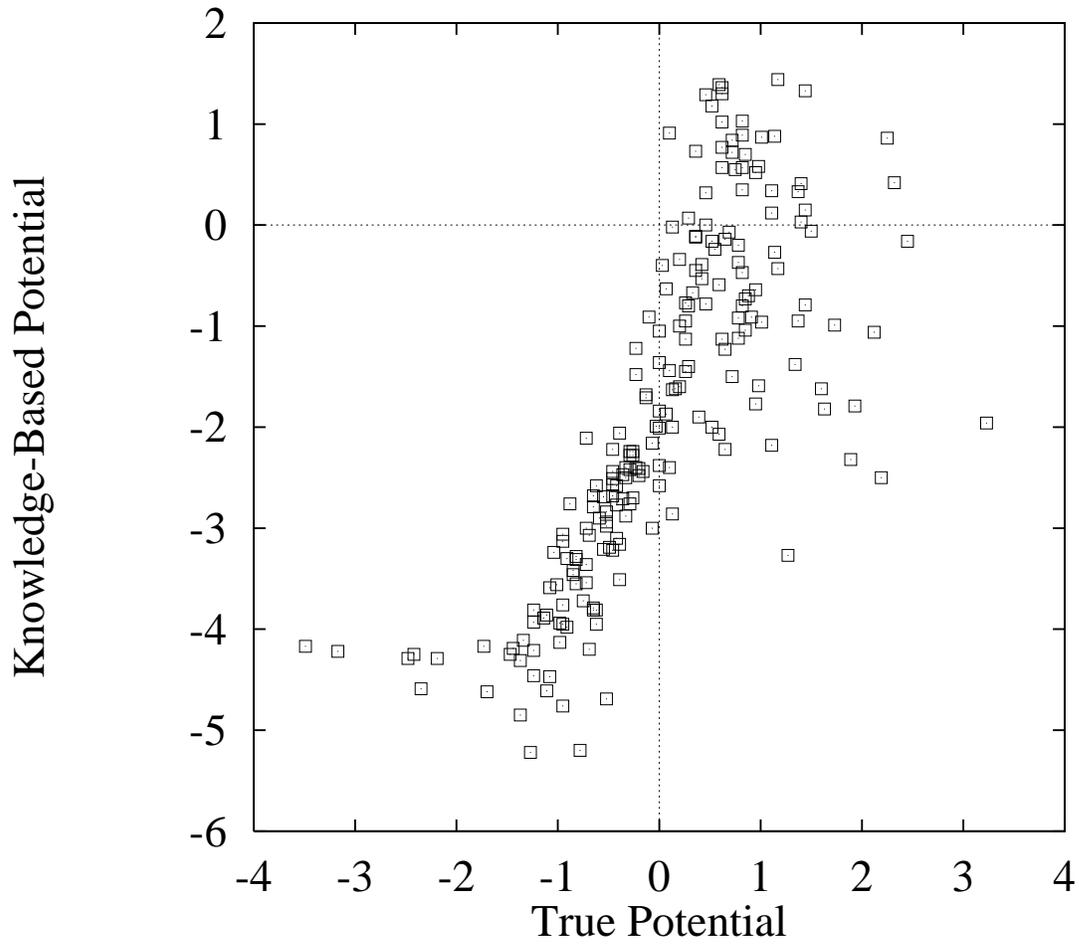}
\vspace{0.5in}
\caption{Potential obtained by statistical 
knowledge based technique vs ``true'' potential for the lattice model.}
\label{fig:mjlike_vs_true}
\end{figure}

All tested techniques are also quite efficient in reconstruction of
the ``true'' potential exhibiting, however, different patterns of
distortion of the original potential. Both optimization techniques
tend to underestimate repulsive interactions (see Figure
\ref{fig:derived_vs_true.lattice}). Knowledge-based techniques, in
contrast, provide good estimates of energies of repulsive
interactions, suffering from underestimation of attractive interactions
(see Figure \ref{fig:mjlike_vs_true}). Attractive interactions are
responsible for stabilization of the native conformation and
underestimation of attractive interactions leads to the observed
(Fig. \ref{fig:z_vs_id2}) increase in $Z$ score for knowledge-based
potentials.

Another deformation of the ``true'' potential by MJ technique is
that it yields
strong non-specific attraction between residues, which
is seen as low negative average interaction between residues ($\langle
U_{MJ} \rangle = -1.07 $ when $\sigma (U) $ set to 1).  This
non-specific attraction favors more compact conformation irrespective
to amino acid sequence. This effect can mislead {\it ab initio}
folding and fold recognition. The origin of this non-specific
attraction is in residue-solvent interactions taken into account by MJ
procedure.  Estimate of the number of solvent-solvent interactions is
responsible for the non-specific attraction. \marginpar{Fig.10}

Although all derivation procedures reconstruct the ``true'' potential
with systematic deviations, all potentials are able to provide
large enough energy gap for well designed model sequences.

\section*{DISCUSSION}
In this work we proposed and tested a novel systematic 
approach to the long-standing problem of how to find 
the correct potential for protein folding. 

In contrast to widely-used knowledge-based statistical 
technique, which relies on
hardly justifiable assumption of
Boltzmann statistics, we use optimization in space of parameters 
to search for a
potential which maximizes stability of all  native proteins
in the dataset.

The procedure was tested using the  ``ideal'' model where sequences
were designed with some known, ``true'' potential and 
the recovered potential turned out to be quite close
to the ``true'' one. The key feature of 
``ideal'' models (both lattice and off-lattice)
is that the form of the energy function (two-body contact Hamiltonian) 
is ``exact'', and the goal of the parameter
search  is to determine 210 numbers - parameters
of this Hamiltonian. We showed that our procedure recovers
the parameters reliably and uniquely.  It is important to 
note that it is not crucial for our method 
that sequences in the database
are well-designed: in fact derivation of potentials 
using the database
of weakly designed sequences (i.e. having relatively high Z-score)
yielded the potentials which were similarly quite close to the
``true'' potential. This is in contrast to the control 
case of assigning randomly shuffled sequences
to structures: for them our procedure did not
converge to any meaningful potential. In this case addition of any
new ``protein'' (in fact a structure with a random sequence assigned
to it) changed the potential dramatically, consistent with the notion
that there is no potential which delivers low energy to all 
structure-random sequence pairs. In contrast, even for weakly designed
sequences there is the potential for which these sequences have
low (but perhaps not the lowest) energy in their corresponding native 
conformations, and such potential is readily recovered by our optimization
technique.

The method has internal controls of self-consistency. First is that
the optimization procedure in parameter space converges
rapidly and at all algorithmic temperatures to unique solution (no multiple-minima problem in space
of parameters). This suggests
that the obtained solution delivers 
global minimum of $Z$ scores for studied 
proteins: no other potential can provide, in average, lower
$Z$ scores for the same structures and sequences in the same model.  

Another important test of self-consistency of the proposed method
is convergence of potentials when the database 
size grows. This clearly points
out that the problem is not undertermined as well as it indicates
clearly that there indeed exists a potential with which all structures
have low energy. This criterion is especially important and useful
when we consider more complicated, than pairwise contact, energy
functions (see below).

The ``ideal'' models provide ideal opportunity to compare our new
method with other approaches,
in particular with the methods based on quasichemical approximation.
 Comparison of the ``true'' potential
with the ones derived by Miyazawa and Jernigan (1996) and Hinds and Levitt (1995) methods (including most demanding 
{\it ab initio} folding
tests)  shows that procedures based on the quasichemical approximation 
can extract potential with impressive accuracy.
This conclusion is in contrast with the assertion of 
 Thomas and Dill (1996) who also tested MJ procedure
using lattice model and argued that the extracted potential
is not an accurate approximation of the ``true'' potential.
We believe that the most important 
criterion of success of extracted potential
is how it performs in {\it ab initio} folding or threading  tests.
Thomas and Dill's test is similar in spirit
to threading because they addressed the issue of how often
the global energy minimum structure remains such with extracted
potentials, judged by exhaustive enumeration of conformations.
They considered all sequences (having unique native state)
for 14-mers and 16-mers with random sequences. However, the
native conformations of random sequences
(as well as other sequences having no or minimal energy gap)
are extremely unstable with respect 
to any uncertainties in potentials
(Bryngelson, 1993).
This is in contrast with folding sequences (with energy gaps) 
which were shown to be much more robust 
with respect to uncertainties in potentials (Pande et al, 1995).
Unfortunately, there are no sequences in HP model which have 
energy gaps (Thomas {\em et al}, 1995), and one cannot even design such sequences  in HP model.

Therefore 
we believe that the major reason of the conclusion 
made by Thomas and Dill (1996) is that they used the model
where native conformation is 
unstable with respect to even minor uncertainties
in potentials. It is important to note also that 
models where 
sequences do not have energy gap 
are equally unstable with respect 
to point mutations (Shakhnovich and Gutin, 1991). 
The remarkable stability
of proteins with respect to many point mutations is another
strong  evidence
that real proteins should have 
a pronounced energy gap, a property absent in HP models.

Building a set of alternative conformations to compute $Z$ score is an
important part of this work. All results discussed here have been
obtained under the following assumptions regarding the presentation of
alternative conformations: (i) all alternative conformations have the
same compactness as the native conformation; (ii) all contacts are
equally probable and (iii) they are 
statistically independent in the set of
alternative conformations. These assumptions allowed us to compute $Z$
score for a protein without building the set of alternative
conformations explicitly. In fact, in order 
to compute the $Z$ score for pairwise
potential one needs to calculate 
the average frequency of a contact and covariance
of two contacts in the set of alternative conformations. The first
assumption states that the number of contacts in alternative
conformations is the same as in the native one. Assuming compactness
of alternative conformations we eliminate the 
effect of non-specific
attraction/repulsion in the recognition of the native
conformation. Non-specific attraction introduced into potential favors
most compact conformations irrespective of amino acid sequence, which
can give rise to  false positives: Very compact 
low energy conformations for any protein in the fold recognition
test. One should be careful about non-specific term in a potential as
it can substantially affect the results of {\it ab initio} folding or
fold recognition. On the other hand, these nonspecific terms can be readily
eliminated by shifting the parameters by a given value (Shakhnovich, 1994;
Gutin {\em et al}, 1995).

By assuming equal probability for all contacts we neglect slight
prevalence of contacts between amino acids close to each other along
the polypeptide chain, which exists even in random coil. However, since
alternative conformations have the same compactness as the native one
local contacts
are not expected to dominate in these conformations
(Abkevich {\em et al}, 1995). 
Different probabilities of local and non-local contacts
can be taken into account by assigning higher probabilities to local
contacts in the set of unfolded conformations 
used in calculation of $Z$-score.

Our assumption of independent contacts is strictly valid only for
point-size non-connected objects. Chain connectivity enforces positive correlation between contacts $i,j$ and $i,j+l$
for small $l=1,2..$.  On the other hand, excluded volume
 of amino acids 
leads to anti-correlation between contacts $i,j$ and $i,k$ since amino
acid $i$ can have only a limited number of contacts due to 
excluded volume
interactions. Several other factors can contribute to 
correlation between
contacts in the opposite ways and 
the final outcome of these effects
has yet to be understood.

The set of alternative conformations built in this way  turned out to
be adequate for estimating the energy gap, in our model, since lattice
proteins which have low enough $Z$ scores are able to fold fast to
their native conformations.

In general, while deriving a potential for a particular task and sampling
procedure (fold recognition, design of an inhibitor, {\it ab initio}
folding etc) one has to construct a set of alternative conformations
which which will be used as decoys during the sampling.

Alternative conformations used in this work correspond more closely
(though not exactly) to sampling by
folding under the condition of average attractive interaction
between aminoacids, while fold recognition is likely to 
have a different set of decoys. This set  of decoys 
should be used in our procedure 
of derivation of potentials for fold recognition.
It can be implemented by explicit generation of alternative
conformations for a given protein by threading its sequence through
other proteins structures of the database.  Frequency of contacts and
contact correlations computed for alternative conformations built in
this way are to be used for computing $Z$ scores and derivation of
potential. While derived, the potential will provide the highest
possible $Z$ score for fold recognition. This work is currently in progress.


Now we turn to the discussion of the 
results obtained for real proteins.
First of all we see that pairwise 
contact approximation is not meaningless for
real proteins, i.e. certain aspects 
of their energetics are captured by that model.
The clear evidence for that is 
that our procedure converges to a unique
potential, and the $Z$-scores 
of proteins with that potential
are considerably lower than for 
randomly shuffled sequences. This suggests
that such a simplified Hamiltonian 
still carries  some 
``signal''.  In this sense
the derived two-body potentials 
are useful since they are able
to discriminate between  native 
conformation and decoys, when the number of decoys is not too large. 

However,  the $Z$-scores obtained for proteins within the
pairwise contact Hamiltonian approximation are not sufficiently
low to provide high
stability (or large energy gap) 
for all proteins simultaneously. Hence all knowledge-based
potentials can  have only
limited success in folding or recognition of the native fold among
alternative conformations. This result can help in  
understanding of the origin of 
problems arising with various structure prediction
techniques. Our results suggest that limited success
in folding simulations in the simple model with pairwise potentials
may be due not to incorrect ``potentials'' (i.e. 210 numbers) 
but rather
due to the deficiency of the model itself, and no other potentials
within the same model of pairwise contact interactions can provide better
results uniformly for numerous tested proteins
(of course there can be successes with ``potential''
which are optimized to fold just one protein
(Hao and Sheraga, 1996); however, as our analysis shows,
such potential (speaking in our terms, derived by optimization 
from the database of
one protein) will fail when used to fold another protein.


Several models have been suggested for protein folding which vary in
accuracy of structure representation and in complexity of the energy
function. What is the optimal number of parameters of the energy
function? How does the number of parameters affect the results of 
the procedure to extract  potentials ?
To address these questions we developed convergence test which allows
to estimate stability of the obtained potential with respect to the
dataset of proteins used. This test indicates whether the number of
parameters used to maximize the energy gap (210 in the case on contact
pairwise potential) is large enough to provide the gap for all
proteins simultaneously and is small enough not to overfit the data
and adopt any random sequence to a protein structure in the
database. Our results
indicate that 210 parameters of contact pairwise potential are not too
many (potential converges as the size of the database increases), but
the model itself is not sufficiently realistic 
to provide the large gap for real proteins. 
More accurate presentation
of energy function, (possibly including 
local conformational preferences, distance dependent interactions,
multibody interactions
etc.) is likely to be necessary to achieve better
discrimination between the native structure and decoys.
The presented method allows to assess systematically 
the validity of
different models and therefore can serve as a powerful tool
for the search of the most adequate model for protein folding. 

\section*{Electronic address to get the parameter set}
The parameter set is available from our anonymous ftp-site 
paradox.harvard.edu; file Euv.dat  in the directory /pub/leonid

\section*{Acknowledgements} 

We are grateful to Alexander Gutin and Victor
Abkevich for many fruitful discussions. We thank Richard Goldstein for
explaning to us their averaging procedure. This work was funded
by Packard Foundation.

\pagebreak

\section*{References}

\noindent
Abkevich,V.I.,  Gutin,A.M~ \& Shakhnovich, E.I. (1994)
Free energy landscape for protein folding kinetics. intermediates,
  traps and multiple pathways in theory and lattice model simulations. {\em
  J.Chem.Phys} {\bf 101}, 6052--6062.

\noindent
Abkevich,V.I.,  Gutin,A.M \& Shakhnovich,E.I (1995)
Impact of local and non-local interactions on thermodynamics and
  kinetics of protein folding.  {\em J. Mol.Biol.} {\bf
  252}, 460--471.

\noindent
Bowie, J. U., Luthy, R. \& Eisenberg, D. (1991). A method to
identify protein sequences that fold into a known three-dimensional
structure. {\em Science} {\bf 253}, 164-170.

\noindent
Bryant, S. H. \& Lawrence, C. E. (1993). An empirical energy function
for threading protein sequence through the folding motif. {\em Proteins:
Struct. Funct. Genet.} {\bf 16}, 92-112.

\noindent
Bryngelson, J.D. (1994). When is a potential accurate enough for structure
prediction?  {\em J.Chem.Phys.} {\bf 100}, 6038-6045.

\noindent
Covell, D.G. (1994) Low resolution models of polypeptide chain collapse.
{\em J. Mol.Biol.} {\bf 25}, 1032-1043 

\noindent
DeWitte, R \& Shakhnovich, E. (1994) Pseudodihedrals: simplified
protein backbone representation with knowledge-based energy.
{\em Protein Science}, {\bf 3}, 1570-1581

Elofson, A. Le Grand, S. \& Eisenberg, D (1995) Local moves: An 
efficient algorithm for simulation of protein folding. {\em Proteins: Struct. Funct. Genet.}
{\bf 23}, 73-82

\noindent
Finkelstein,A.V.  \& Reva, B.A. (1991)
Search for the most stable folds of protein chains. {\em Nature.}
{\bf 351}, 497-499.

\noindent
Finkelstein, A. V., Gutun, A. M. \& Badretdinov, A. Ya. (1993). Why are
the same protein folds used to perform different functions? {\em FEBS
Letters} {\bf 325}, 23-28.

\noindent
Finkelstein, A. V., Gutin, A. M. \& Badretdinov,
A. Ya. (1995). Why do protein
architectures have Boltzmann-like statistics? {\em Proteins: Struct. Funct. Genet.} {\bf 23}, 142-149. 

\noindent
Godzik,A; Kolinski, A \& Skolnick, J. (1995) Are proteins ideal mixtures of amino acids? Analysis of energy parameter sets. {\em Protein Science} {\bf 4}, 2101-2117.

\noindent
Goldstein,R.,  Luthey-Schulten,Z.A.  \& Wolynes,P. (1992).
Optimal protein-folding codes from spin-glass theory. {\em Proc.
  Natl. Acad. Sci. USA} {\bf 89}, 4918--4922.

\noindent
Gutin,A.M.,  Abkevich,V.I. \& Shakhnovich,E.I.  (1995)
Evolution-like selection of fast-folding model proteins.  
{\em Proc   Natl. Acad. Sci. USA} {\bf 92}, 1282--1286.

\noindent
Hao, M. \& Scheraga, H. (1996) How optimization of potential function
affects protein folding. {\em Proc. Natl. Acad. Sci. USA}  {\bf 93}, 4984-4989.

\noindent
Hinds, D. A. \& Levitt, M. (1994) Exploring conformational space with a
simple lattice model for protein structure. {\em J. Mol. Biol}. {\bf 243},
668-82.

\noindent
Jones, D. T., Taylor, W. R. \& Thornton, J. M. (1992). A new approach
to protein fold recognition. {\em Nature} {\bf 358}, 86-89.

\noindent
Kocher, J. P., Rooman, M. J. \& Wodak, S. J. (1994). Factors
influencing the ability of knowledge-based potentials to identify
native sequence-structure matches. {\em J. Mol. Biol}. {\bf 235}, 1598-1613.

\noindent
Kolaskar, A. S. \& Prashanth, D. (1979). Empirical torsional potential
functions from protein structure data. {\em Int. J. Peptide Protein
Res.} {\bf 14}, 88-98.

\noindent
Kolinski,A.  \& Skolnick,J. (1993) A general method for the prediction of three dimensional structure and folding pathway of globular proteins: Application to designed helical proteins. {\em J.Chem.Phys.}, {\bf 98}, 7420-7433.

\noindent
Kolinski, A. \& Skolnick, J. (1994). Monte Carlo simulations of
protein folding. I. Lattice model and interaction scheme. {\em Proteins:
Struct. Funct. Genet.} {\bf 18}, 338-352.

\noindent
Lattman E.E, ed (1995) 
Protein Structure prediction issue, {\em Proteins:Struct. Funct. Genet.}
{\bf 23} 295-462.

\noindent
Levitt, M (1976). A simplified representation of protein conformation
for rapid simulation of protein folding. {\em J.Mol.Biol}. {\bf 104}, 59-107.

\noindent
Lemer, C., Rooman, M. \& Wodak, S. (1995) Protein structure prediction
by threading methods: evaluation of current techniques {\em Proteins:Struct. Funct. Genet.}
{\bf 23} 321-332.
 
\noindent
Maiorov, V. N. \& Crippen, G. M. (1992) Contact potential that
recognizes the correct folding of globular proteins. {\em J.Mol.Biol.} {\bf 227.}
876-88.

\noindent
Mirny, L \& Domany, E (1996) Protein fold recognition and dynamics in
the space of contact maps. {\em Proteins:Struct. Funct. Genet.}, in press

\noindent
Miyazawa, S. \& Jernigan, R. L. (1985). Estimation of effective
interresidue contact energies from protein crystal structures:
quasi-chemical approximation. {\em Macromolecules}, {\bf 18}, 534-552.

\noindent
Miyazawa,S. \& Jernigan,R. (1996) Residue - Residue 
Potentials with a Favorable Contact 
Pair Term and an Unfavorable High 
Packing Density Term, for Simulation 
and Threading.  {\em J.Mol.Biol.} {\bf 256}, 623-644

\noindent
Nishikawa, K. \& Matsuo, Y. (1993). Development of pseudoenergy
potentials for assessing protein 3-D-1- D compatibility and detecting
weak homologies. {\em Protein Eng}. {\bf 6}, 811-820.

\noindent
Pande, V., Grosberg, A. \& Tanaka, T. (1995), How Accurate Must  
Potentials Be for Successful Modeling of Protein Folding? 
{\em J.Chem.Phys.} {\bf 103}, 9482-9491.

\noindent
Rooman, M., Kocher, J.A. \& Wodak, S (1992) Extracting information
on folding from aminoacid sequence: Accurate predictions for protein regions
with preferred conformation in the absence of tertiary interactions.
{\em Biochemistry}, {\bf 32}, 10226-10238

\noindent
Sali,A. Shakhnovich, E.I. \& Karplus,M. (1994a) Kinetics of protein folding.
A lattice model study of the requirements for folding to the native state.
{\em Journ. Mol. Biol.,}, {\bf 235}, 1614--1636.

\noindent
Shakhnovich, E.I. \& Gutin, A.M. (1991) 
Influence of Point Mutations on Protein Structure. Probability of a Neutral Mutation. {\em J.Theor.Biol.} {\bf 149}, 537-546. 

\noindent
Shakhnovich,E.I., Farztdinov,G.M., Gutin,A.M \& Karplus,M. (1991)
Protein folding bottlenecks: a lattice Monte Carlo simulation.
{\em Phys.Rev.Lett.}, {\bf 67}, 1665-1667.

\noindent
Shakhnovich,E.I. \& Gutin,A.M. (1993a)
 Engineering of stable and fast-folding sequences of model proteins.
  {\em Proc.Natl. Acad. Sci. USA} {\bf 90}, 7195-7198.

\noindent
Shakhnovich,E.I \& Gutin,A.M. (1993b)  
A novel approach to design of stable proteins. {\em Protein
  Engineering} {\bf 6}, 793--800.

\noindent
Shakhnovich,E.I. (1994)  Proteins with selected sequences 
fold to their unique native
  conformation {\em Phys.Rev.Lett.} {\bf 72}, 3907--3910.

\noindent
Sippl, M. J. (1990). Calculation of conformational ensembles from
potentials of mean force. An approach to the knowledge-based
prediction of local structures in globular
proteins. {\em J. Mol. Biol}. {\bf 213}, 859-883.

\noindent
Skolnick, J. \& Kolinski, A. (1990). Simulations of the folding of a
globular protein. {\em Science} {\bf 250}, 1121-1125.

\noindent
Tanaka, S. \& Scheraga, H. A. (1976). Medium- and long-range
interaction parameters between amino acids for predicting
three-dimensional structures of proteins. {\em Macromolecules} {\bf 9}, 945-950.

\noindent
Thomas, P. \& Dill, K. (1996) Statistical potentials extracted from
protein structures: How accurate are they?  {\em J. Mol. Biol.} {\bf 257}, 457-469.

\noindent
Ueda, Y., Taketomi,H. \& Go, N. (1978) Studies of protein folding, unfolding and fluctuations by computer simulations. II A three-dimensional lattice model of lysozyme. {\em Biopolymers}, {\bf 17}, 1531-1548.

\noindent
Vasques, M., Nemethy, G. \& Scheraga, H. (1994)
Conformational energy calculations on polypeptides and proteins.
{\em Chem.Rev.} {\bf 94}, 2183-2239.

\noindent
Wilson, C. \& Doniach, S. (1989). A computer model to dynamically
simulate protein folding: studies with crambin. {\em Proteins:
Struct. Funct. Genet.} {\bf 6}, 193-209.

\noindent
Wodak, S and Rooman, M. (1993) Generating and testing protein folds.
{\em Curr. Opin. Struct. Biol.} {\bf 3} 247-259.

\noindent
Yue,K.,  Fiedig,K., Thomas, P.  Chan, H.S., Shakhnovich, E.I. \& Dill, K.A. 
(1995) A test of lattice protein folding algorithms. 
{\em Proc. Natl. Acad. Sci. USA}
{\bf 92}, 325--329.

\pagebreak

\begin{table}
\begin{tabular}{l r r r l }
\bf Potential      &          $<Z>$                & \bf Correlation         & \bf Fraction of proteins & Ref.  \\
                   &                             & \bf with true potential & \bf able to fold   &       \\
\hline
``True'' potential &          7.68               &        1.00             &    100 \%     &       \\
this work          &          8.61               &        0.83 (0.82)       &     96 \%     &       \\ 
Goldstein {\em et al} &          8.45               &        0.78 (0.71)       &     94 \%     &       \\
Hinds\&Levitt      &          7.18               &        0.86 (0.84)       &     99 \%     &       \\
Miyazawa\&Jernigan &          7.09               &        0.75 (0.68)       &     95 \%     &       \\
\end{tabular}
\label{tab:kb_wg_ms}
\vspace{0.5in}
\caption{Comparison of different procedures for derivation of potential.
Correlations are computed for potentials obtained using all 200 proteins. Correlation shown in brackets are for potentials obtained using 100 proteins}
\end{table}

\newpage

\pagebreak

\pagebreak

\pagebreak

\pagebreak

\pagebreak

\pagebreak

\pagebreak

\pagebreak

\pagebreak

\end{document}